\documentclass[a4paper, final, 12pt]{article}
\usepackage{amsmath, amssymb, latexsym, amscd, amsthm,amsfonts,amstext}
\usepackage[mathscr]{eucal}
\usepackage{graphicx}
\usepackage{subfig}
\usepackage{float}
\usepackage{color}
\usepackage{hyperref}
\usepackage[utf8]{inputenc}
\usepackage[english]{babel}

\providecommand{\func}[1]{\operatorname{#1}}

\numberwithin{equation}{section}
\begin{document}

\title{Simultaneous Reconstructions of $\left( x,t\right) -$Dependent
Coefficients and Initial Conditions of Parabolic Equations }
\author{Michael V. Klibanov \thanks{%
Email: \texttt{mklibanv@charlotte.edu}; supported by the US National Science
Foundation grant DMS 2436227.} \\
{\small Department of Mathematics and Statistics }\\
{\small University of North Carolina at Charlotte, Charlotte, NC 28223, USA }%
}
\date{}
\maketitle

\begin{abstract}
Let $x\in \mathbb{R}^{n}$ be the vector of spatial variables and $t>0$ be
the time variable. This is the first publication, in which an inverse
problem of the simultaneous reconstruction of an unknown $\left( x,t\right) -
$dependent coefficient and an unknown initial condition in a general linear
parabolic equation of the second order is considered. The input data depend
on $\left( n+1\right) $ variables and are, therefore, formally determined
ones. Stability estimates and uniqueness theorem are obtained for this
inverse problem. As a by-product, logarithmic stability estimates for
initial conditions of parabolic equations and inequalities with time
reversed data are obtained for the first time.

The inverse problem of this paper has two potential applications. The first
one is in forecasting of public opinions in the framework of the Mean Field
Games theory. The second one is in tracking spatiotemporal outbreaks of
epidemics.

.
\end{abstract}

\tableofcontents

\markright{Simultaneous Reconstruction}



\newpage

\section{Introduction}

\label{sec:1}

Let $T>0$ be a number. Below $x\in \mathbb{R}^{n}$ denotes the vector of
spatial variables and $t\in \left( 0,T\right) $ denotes the time variable.
All functions considered below are real valued ones.

All past publications about coefficient inverse problems for parabolic
equations of the second order assume that the unknown coefficients depend
either only on $x$ or only on $t$, and, in addition, the initial condition
is known, see some details in this section below. Unlike those, this is the
first publication, which considers an inverse problem of a simultaneous
reconstruction of two unknown functions in a parabolic equation: an $\left(
x,t\right) -$dependent coefficient and the initial condition. An applied
motivation of this work is described below. Our input data are formally
determined ones, since the number $m$ of free variables in our input data
equals the number $n+1$ of free variables in both the unknown coefficient
and the unknown initial condition, $m=n+1$. Hence, our input data are
formally determined ones.

Two main new results of this paper are:

\begin{enumerate}
\item Two H\"{o}lder stability estimates for the unknown coefficient.

\item Logarithmic stability estimate for the initial condition.
\end{enumerate}

Uniqueness for both that coefficient and the initial condition follows
immediately from one of those H\"{o}lder stability estimates. Since the
solution of that parabolic equation depends nonlinearly on its coefficients,
then this inverse problem is nonlinear.\ We consider two cases of the input
data for our inverse problem. In the first case those are the lateral Cauchy
data at a part of the boundary of the domain of interest. We obtain in this
case a H\"{o}lder stability estimate for the unknown coefficient and
uniqueness for both that coefficient and initial condition. In the second
case the input data are Cauchy data at the entire lateral boundary. We
obtain in the latter case the second H\"{o}lder stability estimate for the
unknown coefficient as well as logarithmic stability estimate for the
unknown initial condition.

The input data in both above cases depend on $n+1$ variables $\left(
x,t,x_{0}\right) $, so as both the unknown coefficient and the unknown
initial condition. Here, the point $\left( x,t\right) $ belongs to the
corresponding laterally oriented hypersurface and $x_{0}$ is a 1-D
parameter, which runs along an interval of a straight line. An applied
interpretation of $x_{0}$ is that it is a running point source of
disturbances, such as, e.g. epidemic outbreaks, see the paragraph after the
next one.

As a by-product, logarithmic stability estimates for the initial conditions
of parabolic equations and inequalities with time reversed data are obtained
here for the first time. More precisely, we derive an estimate of the
initial condition of either a parabolic equation or, more generally,
parabolic inequality with the reversed time. Some estimates of integral
norms over the time cylinders of solutions of such equations and
inequalities are known, see, e.g. \cite{KY}, \cite[Theorem 1 in \S 2 of
chapter 4]{LRS}. However, those time cylinders are located in $\left\{
t>\varepsilon >0\right\} $ for some small numbers $\varepsilon ,$ i.e. the
hyperplane $\left\{ t=0\right\} $ is not \textquotedblleft reached". We
address here a far more delicate question about estimates of initial
conditions at $\left\{ t=0\right\} .$ The key novel element, which allows us
to do so, is the presence of the non-negative term $\left( 1/3\right)
\lambda e^{2}\left\Vert u\left( x,0\right) \right\Vert _{L_{2}\left( \Omega
\right) }^{2}$ in the Carleman estimate of Theorem 8.3 in subsection 8.3.
This term is not present in previously known Carleman estimates for
parabolic operators with the reversed time.

The applied motivation of the author is generated by his first hand recent
experience of \cite{MFGforecasting}. The paper \cite{MFGforecasting} is
concerned with the epidemiology. In that paper the real data of public
opinions are treated with the goal of forecasting these opinions. These data
were collected during the COVD-19 pandemic. The forecasting process is
modeled in \cite{MFGforecasting} via the so-called Mean Field Games System
(MFGS). The MFGS consists of two coupled nonlinear parabolic equations with
two opposite directions of time, see, e.g. books \cite{A,KLMFG} about this
system. It is clear from those real data that true coefficients of the MFGS
governing this process depend on both $x$ and $t$. In addition, initial
conditions for that system were actually unknown to us, see section 5 in 
\cite{MFGforecasting}.

It was demonstrated in a number of computational examples of \cite%
{MFGforecasting} that the MFGS indeed governs that process, see, e.g. Figure
3, the last Figure A.1 and other similar figures in \cite{MFGforecasting}.
Based on the results of \cite{MFGforecasting}, the author concludes that it
is quite desirable for applications to study inverse problems like the one
in this paper. One of potential future applied outcomes of the current paper
might be a better accuracy of results for real data, such as, e.g. the ones
of \cite{MFGforecasting}. Even though the inverse problem of this paper is
not concerned with the MFGS, still the results presented below might pave
the way towards the achievement of a better accuracy of results presented in 
\cite{MFGforecasting}. The second potential future applied outcome of the
current paper is in tracking spatiotemporal distributions of certain
parameters and initial conditions of epidemics, see, e.g. \cite%
{Klibepid1,Klibepid2} for some relevant publications. Initial conditions in
this case provide an important information about outbreaks of epidemics.

As stated above past publications about inverse problems for parabolic
equations with unknown coefficients assume that those coefficients depend
either only on $x$ or only on $t,$ and that the initial condition is known,
see, e.g. \cite{Bell,BukhKlib,Hamr,Isakov,Klib92,Ksurvey,KL,KLMFG,Liu}. We
also refer to the paper \cite{Ali}, where uniqueness is proven for the case $%
n=1$ with the unknown $x-$dependent coefficient and the unknown initial
condition. Recently a group of authors has posted an intriguing preprint 
\cite{ChenLiu}, in which they have proven uniqueness of simultaneous
reconstructions of $x-$dependent coefficients and initial conditions of
hyperbolic, parabolic and Schr\"{o}dinger equations from the lateral Cauchy
data. Those unknown coefficients and initial conditions are assumed to have
a form of separation of certain variables in \cite{ChenLiu}. This assumption
is not imposed in the current paper. The case of $\left( x,t\right) -$%
dependent unknown coefficients is not considered in \cite{ChenLiu}.

To arrange the dependence on $\left( n+1\right) $ variables of the input
data, we assume here that the unknown initial condition $f$ \ has the form $%
f\left( x,x_{0}\right) ,$ where $x_{0}$ runs along an interval of a straight
line. In fact, $x_{0}$ is an analog of a point source. For example,%
\begin{equation*}
f\left( x,x_{0}\right) =f_{1}\left( x\right) \exp \left( -\frac{\left(
x_{1}-x_{0}\right) ^{2}+x_{2}^{2}+...+x_{n}^{2}}{\theta ^{2}}\right) ,
\end{equation*}%
where the function $f_{1}\left( x\right) $ is unknown, $\theta >0$ is a
number, and the analog of the point source is $\left( x_{0},0,...,0\right) ,$
$x_{0}\in \left( a,b\right) ,$ where $\left( a,b\right) \subset \mathbb{R}$
is a certain finite interval.

Due to an obvious substantial challenge of the problem of this paper, it is
natural to impose a simplifying assumption. Thus, we assume that a certain
function generated by the solution of the governing parabolic PDE admits an
expansion in a truncated Fourier-like series with respect to a special
orthonormal basis in $L_{2}\left( a,b\right) $. This basis was originally
proposed by the author in \cite{Klib2017} for purely numerical rather than
analytical purposes, also, see \cite[section 6.2.3]{KL}. Functions of this
basis depend only on the parameter $x_{0}\in \left( a,b\right) .$ That
truncated Fourier-like series has $N$ terms, where $N>1$ is a fixed
arbitrary integer.

To achieve the goal stated in the above item 1, we substantially modify the
method, which was first introduced in \cite{BukhKlib}. This method is based
on Carleman estimates. The idea of \cite{BukhKlib} was explored in many
publications. In this regard we cite here some samples of those publications
and refer to the references cited therein \cite%
{Bell,Hamr,Isakov,Klib92,Ksurvey,KL,KLMFG,Liu}. It is also well known that
Carleman estimates are used for proofs of unique continuation results, see,
e.g. remarkable results of Ionescu and Klainerman \cite{IK1,IK2}.

\section{Statement of Inverse Problem 1}

\label{sec:2}

Below it is convenient sometimes to re-denote points $x=\left(
x_{1},x_{2},...,x_{n}\right) \in \mathbb{R}^{n}$ as $x=\left( x_{1},y\right)
=\left( x_{1},y_{1},...,y_{n-1}\right) \in \mathbb{R}^{n}.$ The true meaning
of such notations is always clear from the context below. Let $\Omega
\subset \mathbb{R}^{n}$ be a bounded domain with its boundary $\partial
\Omega .$ Let $\Gamma \subset \partial \Omega $ be a part of that boundary.
We assume that%
\begin{equation}
\left. 
\begin{array}{c}
\Omega \subset \left\{ x_{1}>0\right\} , \\ 
\Gamma =\left\{ x=\left( x_{1},y\right) :x_{1}=0,\text{ }\left\vert
y\right\vert <Y,i=1,...,n-1\right\} ,%
\end{array}%
\right.  \label{2.1}
\end{equation}%
where $Y>0$ is a certain number. Denote%
\begin{equation}
Q_{T}=\Omega \times \left( 0,T\right) ,\text{ }S_{T}=\partial \Omega \times
\left( 0,T\right) ,\text{ }\Gamma _{T}=\Gamma \times \left( 0,T\right) .
\label{2.2}
\end{equation}%
Below $\alpha =\left( \alpha _{1},...,\alpha _{n}\right) $ is the multiindex
with non-negative integer coordinates. Denote 
\begin{equation*}
D_{x}^{\alpha }=\frac{\partial ^{\left\vert \alpha \right\vert }}{\partial
^{\alpha _{n}}x_{n}...\partial ^{\alpha _{1}}x_{1}},\text{ }\left\vert
\alpha \right\vert =\sum\limits_{i=1}^{n}\alpha _{i}.
\end{equation*}

Consider a linear elliptic operator $L$ of the second order in $Q_{T}$,%
\begin{equation}
Lu=\sum\limits_{\left\vert \alpha \right\vert \leq 2}a_{\alpha }\left(
x,t\right) D_{x}^{\alpha }u,\text{ }  \label{2.3}
\end{equation}%
\begin{equation}
L_{0}u=\sum\limits_{\left\vert \alpha \right\vert =2}a_{\alpha }\left(
x,t\right) D_{x}^{\alpha }u,  \label{2.30}
\end{equation}%
where $L_{0}$ is the principal part of the operator $L$. We assume that%
\begin{equation}
\left. a_{\alpha }\in C^{1}\left( \overline{Q}_{T}\right) ,\text{ }%
\left\vert \alpha \right\vert =2;\text{ }\max_{\left\vert \alpha \right\vert
=2}\left\Vert a_{\alpha }\right\Vert _{C^{1}\left( \overline{Q}_{T}\right)
}\leq B,\text{ }\right.  \label{2.6}
\end{equation}%
\begin{equation}
a_{\alpha }\in C\left( \overline{Q}_{T}\right) ,\text{ }\left\vert \alpha
\right\vert \leq 1,\text{ }\max_{\left\vert \alpha \right\vert \leq
1}\left\Vert a_{\alpha }\right\Vert _{C\left( \overline{Q}_{T}\right) }\leq
B,  \label{2.60}
\end{equation}%
where $B>0$ is a number. The ellipticity of the operator $L$ means that
there exists a number $\mu >0$ such that%
\begin{equation}
\left. 
\begin{array}{c}
\mu \left\vert \xi \right\vert ^{2}\leq \sum\limits_{\left\vert \alpha
\right\vert =2}a_{\alpha }\left( x,t\right) \xi ^{\alpha },\text{ }\forall
\xi \in \mathbb{R}^{n},\text{ }\forall \left( x,t\right) \in Q_{T}, \\ 
\xi ^{\alpha }=\Pi _{i=1}^{n}\xi _{i}^{\alpha _{i}}, \\ 
\text{and the quadratic form }\sum\limits_{\left\vert \alpha \right\vert
=2}a_{\alpha }\left( x,t\right) \xi ^{\alpha }\text{ is symmetric.}%
\end{array}%
\right.  \label{2.5}
\end{equation}%
Let $x_{0}$ be a parameter, 
\begin{equation}
x_{0}\in \left[ a,b\right] \subset \mathbb{R}.  \label{2.7}
\end{equation}%
Taking into account (\ref{2.7}),\emph{\ }we assume that the function $f$ is:%
\begin{equation}
f\left( x,x_{0}\right) ,\partial _{x_{0}}f\left( x,x_{0}\right) \in C\left( 
\overline{\Omega }\right) \times C\left[ a,b\right] .  \label{2.8}
\end{equation}%
Here $C\left( \overline{\Omega }\right) \times C\left[ a,b\right] $ is the
space of functions $g\left( x,x_{0}\right) ,$ which are continuous in $%
\overline{\Omega }\times \left[ a,b\right] ,$ and the norm is defined as 
\begin{equation*}
\left\Vert g\left( x,x_{0}\right) \right\Vert _{C\left( \overline{\Omega }%
\right) \times C\left[ a,b\right] }=\max_{\left( x,x_{0}\right) \in 
\overline{\Omega }\times \left[ a,b\right] }\left\vert g\left(
x,x_{0}\right) \right\vert ,\text{ }\forall g\left( x,x_{0}\right) \in
C\left( \overline{\Omega }\times \left[ a,b\right] \right) .
\end{equation*}%
Other similar spaces below, like $X\times C\left[ a,b\right] ,$\emph{\ }%
where $X$ is a Banach space, are defined similarly, so as norms in them:%
\begin{equation*}
\left\Vert \cdot \right\Vert _{X\times C\left[ a,b\right] }=\max_{x_{0}\in %
\left[ a,b\right] }\left\Vert \cdot \right\Vert _{X}.
\end{equation*}

Let the function 
\begin{equation}
u\left( x,t,x_{0}\right) \in C^{2,1}\left( \overline{Q}_{T}\right) \times C%
\left[ a,b\right]  \label{2.80}
\end{equation}%
be a solution of the following parabolic equation%
\begin{equation}
u_{t}=Lu,\text{ }\left( x,t,x_{0}\right) \in Q_{T}\times \left( a,b\right) .
\label{2.9}
\end{equation}%
Suppose that the initial condition for $u\left( x,t,x_{0}\right) $ is%
\begin{equation}
u\left( x,0,x_{0}\right) =f\left( x,x_{0}\right) ,\text{ }\left(
x,x_{0}\right) \in \Omega \times \left( a,b\right) .  \label{2.10}
\end{equation}%
Let $\partial _{n}u\mid _{\Gamma _{T}}$denotes the normal derivative of the
function $u\left( x,t\right) $ at $\Gamma _{T}.$ Define two functions $%
g_{0}\left( x,t,x_{0}\right) $ and $g_{1}\left( x,t,x_{0}\right) ,$%
\begin{equation}
u\mid _{\Gamma _{T}}=g_{0}\left( x,t,x_{0}\right) ,\text{ }\partial
_{n}u\mid _{\Gamma _{T}}=g_{1}\left( x,t,x_{0}\right) ,\text{ }x_{0}\in
\left( a,b\right) .  \label{2.11}
\end{equation}

The first inverse problem we study in this paper is Inverse Problem 1.

\textbf{Inverse Problem 1 (IP1). }\emph{Assume that conditions (\ref{2.8})
hold for the function }$f\left( x,x_{0}\right) .$\emph{\ Let the function }$%
u\left( x,t,x_{0}\right) $\emph{\ satisfies conditions (\ref{2.80})-(\ref%
{2.11}), where functions }$g_{0}$\emph{\ and }$g_{1}$\emph{\ are known. Let }%
$\alpha _{0},\left\vert \alpha _{0}\right\vert \leq 2$\emph{\ be a fixed
multiindex. Suppose that the coefficient }$a_{\alpha _{0}}\left( x,t\right) $%
\emph{\ of the operator }$L$\emph{\ in (\ref{2.3}) is unknown. Determine the
pair of functions }$\left( a_{\alpha _{0}}\left( x,t\right) ,f\left(
x,x_{0}\right) \right) .$

\section{The Differentiability of the Function $u\left( x,t,x_{0}\right) $
With Respect to $x_{0}$}

\label{sec:3}

The goal of this section is to show that our condition imposed below of the
continuous differentiability with respect to $x_{0}$ of the function $%
u\left( x,t,x_{0}\right) $ in (\ref{2.9})\emph{\ }can be rigorously
guaranteed in some reasonable scenarios.\emph{\ }We need to impose in this
section more stringent than ones in section 2 regularity conditions on $%
\partial \Omega ,$ the function $f$ and the coefficients of the operator $L$.

\textbf{Remark 3.1.} \emph{We note that extra regularity assumptions are
traditionally of a minor concern in the field of inverse problems, see, e.g. 
\cite{Nov,Rom}.}

Below $\beta \in \left( 0,1\right) $ and $C^{k+\beta },C^{k+\beta ,k+\beta
/2}$ are H\"{o}lder spaces \cite{Lad}, where $k\geq 0$ is an integer. Only
in this section, we assume that 
\begin{equation}
\partial \Omega \in C^{4+\beta },\text{ }a_{\alpha }\left( x,t\right) \in
C^{2+\beta }\left( \overline{Q}_{T}\right) ,\text{ }\left\vert \alpha
\right\vert \leq 2.  \label{3.1}
\end{equation}%
As to the initial condition $f\left( x,x_{0}\right) ,$ we impose, again only
in this section, more stringent regularity conditions than ones in (\ref{2.8}%
),%
\begin{equation}
f\left( x,x_{0}\right) ,\partial _{x_{0}}f\left( x,x_{0}\right) \in
C^{4+\beta }\left( \overline{\Omega }\right) \times C\left[ a,b\right] .
\label{3.2}
\end{equation}%
Consider a function $p\left( x,t,x_{0}\right) $ defined on $S_{T}\times %
\left[ a,b\right] $ and assume that \cite[Theorem 5.2 of Chapter 4]{Lad} 
\begin{equation}
p\left( x,t,x_{0}\right) ,\partial _{x_{0}}p\left( x,t,x_{0}\right) \in
C^{4+\beta ,2+\beta /2}\left( \overline{S}_{T}\right) \times C\left[ a,b%
\right] .  \label{3.3}
\end{equation}%
Consider the initial boundary value problem for parabolic equation (\ref{2.9}%
) in $Q_{T}$ with the initial condition (\ref{2.10}) and with the Dirichlet
boundary condition%
\begin{equation}
u\mid _{S_{T}}=p\left( x,t,x_{0}\right) ,\text{ }\left( x,t,x_{0}\right) \in
S_{T}\times \left( a,b\right) .  \label{3.4}
\end{equation}%
In addition, we assume classical the so-called \textquotedblleft
compatibility conditions" \cite[\S 5 of chapter 4]{Lad}.

\textbf{Theorem 3.1. }\emph{Assume that conditions (\ref{2.3})-(\ref{2.8})
and (\ref{3.1})-(\ref{3.3}) hold. In addition, let the compatibility
conditions up to the order 3 are valid. Then for every }$x_{0}\in \left[ a,b%
\right] $\emph{\ there exists unique solution }$u\left( x,t,x_{0}\right) $%
\emph{\ of equation (\ref{2.9}) with initial condition (\ref{2.10}) and
boundary condition (\ref{3.4}) such that }%
\begin{equation}
u\in C^{4+\beta ,2+\beta /2}\left( \overline{Q}_{T}\right) \times C\left[ a,b%
\right] .  \label{3.5}
\end{equation}%
\emph{The function }$u\left( x,t,x_{0}\right) $\emph{\ is continuously
differentiable with respect to }$x_{0}\in \left[ a,b\right] $\emph{\
together with its corresponding }$\left( x,t\right) -$\emph{derivatives, i.e.%
}%
\begin{equation}
u,\text{ }\partial _{x_{0}}u\in C^{4+\beta ,2+\beta /2}\left( \overline{Q}%
_{T}\right) \times C\left[ a,b\right] .  \label{3.6}
\end{equation}%
\emph{Furthermore, there exists a number }$C_{0}>0$\emph{\ depending only on
coefficients of the operator }$L$\emph{, the domain }$Q_{T}$\emph{\ and the
interval }$\left( a,b\right) $\emph{\ such that the following estimates hold}%
\begin{equation}
\left\Vert u\right\Vert _{C^{4+\beta ,2+\beta /2}\left( \overline{Q}%
_{T}\right) \times C\left[ a,b\right] }\leq C_{0}\left( \left\Vert
f\right\Vert _{C^{4+\beta }\left( \overline{\Omega }\right) \times C\left[
a,b\right] }+\left\Vert p\right\Vert _{C^{4+\beta ,2+\beta /2}\left( 
\overline{S}_{T}\right) \times C\left[ a,b\right] }\right) ,  \label{3.7}
\end{equation}%
\begin{equation}
\left. 
\begin{array}{c}
\left\Vert \partial _{x_{0}}u\right\Vert _{C^{4+\beta ,2+\beta /2}\left( 
\overline{Q}_{T}\right) \times C\left[ a,b\right] }\leq \\ 
\leq C_{0}\left( \left\Vert \partial _{x_{0}}f\right\Vert _{C^{4+\beta
}\left( \overline{\Omega }\right) \times C\left[ a,b\right] }+\left\Vert
\partial _{x_{0}}p\right\Vert _{C^{4+\beta ,2+\beta /2}\left( \overline{S}%
_{T}\right) \times C\left[ a,b\right] }\right) .%
\end{array}%
\right.  \label{3.8}
\end{equation}

\textbf{Proof}. Existence and uniqueness of the solution $u\in C^{4+\beta
,2+\beta /2}\left( \overline{Q}_{T}\right) $ for every $x_{0}\in \left[ a,b%
\right] $ of the above initial boundary value problem as well as estimate (%
\ref{3.7}) follow immediately from \cite[Theorem 5.2 of chapter 4]{Lad}, i.e.%
\begin{equation}
\left. 
\begin{array}{c}
\left\Vert u\left( x,t,x_{0}\right) \right\Vert _{C^{4+\beta ,2+\beta
/2}\left( \overline{Q}_{T}\right) }\leq C_{0}\left( \left\Vert f\right\Vert
_{C^{4+\beta }\left( \overline{\Omega }\right) \times C\left[ a,b\right]
}+\left\Vert p\right\Vert _{C^{4+\beta ,2+\beta /2}\left( \overline{S}%
_{T}\right) \times C\left[ a,b\right] }\right) ,\text{ } \\ 
\forall x_{0}\in \left[ a,b\right] .%
\end{array}%
\right.  \label{3.80}
\end{equation}
We now work with the derivative with respect to $x_{0}.$ Fix an arbitrary
point $x_{0}\in \left( a,b\right) .$ Let $h$ be a number such that $%
\left\vert h\right\vert <\xi $ for a sufficiently small $\xi >0.$ Then%
\begin{equation}
\left. 
\begin{array}{c}
\left( f\left( x,x_{0}+h\right) -f\left( x,x_{0}\right) \right) /h-\partial
_{x_{0}}f\left( x,x_{0}\right) =o\left( 1\right) ,\text{ as }h\rightarrow 0,%
\text{ } \\ 
\forall \left( x,x_{0}\right) \in \overline{\Omega }\times \left\{
x_{0}:x_{0}\in \left( a,b\right) ,x_{0}+h\in \left( a,b\right) \right\} .%
\end{array}%
\right.  \label{3.9}
\end{equation}%
And also 
\begin{equation}
\left. 
\begin{array}{c}
\left( p\left( x,t,x_{0}+h\right) -p\left( x,t,x_{0}\right) \right)
/h-\partial _{x_{0}}p\left( x,x_{0}\right) =o\left( 1\right) ,\text{ as }%
h\rightarrow 0, \\ 
\forall \left( x,t,x_{0}\right) \in \overline{S}_{T}\times \left\{
x_{0}:x_{0}\in \left( a,b\right) ,x_{0}+h\in \left[ a,b\right] \right\} .%
\end{array}%
\right.  \label{3.10}
\end{equation}%
Denote%
\begin{equation}
g_{f}\left( x,x_{0},h\right) =\frac{f\left( x,x_{0}+h\right) -f\left(
x,x_{0}\right) }{h},  \label{3.100}
\end{equation}%
\begin{equation}
g_{p}\left( x,t,x_{0},h\right) =\frac{p\left( x,t,x_{0}+h\right) -p\left(
x,t,x_{0}\right) }{h}.  \label{3.101}
\end{equation}%
Consider the solution $\overline{u}\left( x,t,x_{0},h\right) \in C^{4+\beta
,2+\beta /2}\left( \overline{Q}_{T}\right) $ of the following initial
boundary value problem%
\begin{equation}
\left. 
\begin{array}{c}
\overline{u}_{t}=L\overline{u},\text{ }\left( x,t\right) \in Q_{T}, \\ 
\overline{u}\left( x,0,x_{0},h\right) =g_{f}\left( x,x_{0},h\right) ,\text{ }%
x\in \Omega , \\ 
\overline{u}\mid _{S_{T}}=g_{p}\left( x,t,x_{0},h\right) .%
\end{array}%
\right.  \label{3.11}
\end{equation}%
In addition, consider the solution $\widehat{u}\left( x,t,x_{0}\right) \in
C^{4+\beta ,2+\beta /2}\left( \overline{Q}_{T}\right) $ of the following
initial boundary value problem%
\begin{equation}
\left. 
\begin{array}{c}
\widehat{u}_{t}=L\widehat{u},\left( x,t\right) \in Q_{T}, \\ 
\widehat{u}\left( x,0,x_{0}\right) =\partial _{x_{0}}f\left( x,x_{0}\right) ,%
\text{ }x\in \Omega , \\ 
\widehat{u}\mid _{S_{T}}=\partial _{x_{0}}p\left( x,t,x_{0}\right) .%
\end{array}%
\right.  \label{3.12}
\end{equation}%
Subtracting (\ref{3.12}) from (\ref{3.11}) and denoting 
\begin{equation}
U\left( x,t,x_{0}\right) =\overline{v}\left( x,t,x_{0},h\right) -\widehat{v}%
\left( x,t,x_{0}\right) ,  \label{3.13}
\end{equation}%
we obtain%
\begin{equation}
\left. 
\begin{array}{c}
U_{t}=LU,\text{ }\left( x,t\right) \in Q_{T}, \\ 
U\left( x,0,x_{0}\right) =g_{f}\left( x,x_{0},h\right) -\partial
_{x_{0}}f\left( x,x_{0}\right) , \\ 
U\mid _{S_{T}}=g_{p}\left( x,t,x_{0},h\right) -\partial _{x_{0}}p\left(
x,t,x_{0}\right) .%
\end{array}%
\right.  \label{3.14}
\end{equation}%
It follows from (\ref{3.80}), (\ref{3.9})-(\ref{3.101}) and (\ref{3.14})
that 
\begin{equation*}
\left\Vert U\right\Vert _{C^{4+\beta ,2+\beta /2}\left( \overline{Q}%
_{T}\right) }=o\left( 1\right) \text{ as }h\rightarrow 0.
\end{equation*}%
Hence, (\ref{3.100})-(\ref{3.14}) imply 
\begin{equation*}
\widehat{u}\left( x,t,x_{0}\right) =\partial _{x_{0}}u\left(
x,t,x_{0}\right) \in C^{4+\beta ,2+\beta /2}\left( \overline{Q}_{T}\right)
\times C\left[ a,b\right] ,
\end{equation*}%
which proves (\ref{3.6}). Hence, $u\left( x,t,x_{0}\right) \in C^{4+\beta
,2+\beta /2}\left( \overline{Q}_{T}\right) \times C\left[ a,b\right] $ and
also (\ref{3.80}) implies (\ref{3.7}). Estimate (\ref{3.8}) follows from (%
\ref{3.7}) and (\ref{3.12}). $\square $

\section{A Special Orthonormal Basis in $L_{2}\left( a,b\right) $}

\label{sec:4}

We now present a special orthonormal basis in $L_{2}\left( a,b\right) .$
This basis was first constructed in \cite{Klib2017} for purely numerical
purposes, also, see \cite[section 6.2.3]{KL}. Consider the set of functions $%
\{x_{0}^{n}e^{x_{0}}\}_{n=0}^{\infty }\subset L_{2}(a,b).$ These functions
are linearly independent ones and form a complete set in the space $%
L_{2}(a,b)$. Apply the Gram-Schmidt orthonormalization procedure to this
set. Then we obtain an orthonormal basis $\{\Psi _{k}\left( x_{0}\right)
\}_{k=0}^{\infty }$ in $L_{2}(a,b)$. For any $k\geq 0$ the function $\Psi
_{k}(x_{0})$ has the form 
\begin{equation}
\Psi _{k}(x_{0})=P_{k}(x_{0})e^{x_{0}},  \label{4.1}
\end{equation}%
where $P_{k}(x_{0})$ is a polynomial of the degree $k$. The key result for
this basis is Theorem 4.1.

\textbf{Theorem 4.1} (\cite{Klib2017}, \cite[Theorem 6.2.1]{KL}). \emph{Let }%
$\left( ,\right) $\emph{\ be the scalar product in }$L_{2}(a,b)$\emph{.
Consider the numbers }$d_{k,s},$\emph{\ }%
\begin{equation}
d_{k,s}=\left( \Psi _{k}^{\prime },\Psi _{s}\right)
=\int\limits_{-A}^{A}\Psi _{k}^{\prime }\left( x_{0}\right) \Psi _{s}\left(
x_{0}\right) dx_{0}.  \label{4.01}
\end{equation}%
\emph{\ Then}%
\begin{equation}
d_{k,s}=\left\{ 
\begin{array}{c}
1\text{ if }k=s, \\ 
0\text{ if }k<s.%
\end{array}%
\right.  \label{4.2}
\end{equation}%
\emph{Let }$N\geq 1$\emph{\ be an integer. Consider the }$N\times N$\emph{\
matrix }%
\begin{equation}
W_{N}=\left( d_{_{k,s}}\right) _{\left( k,s\right) =\left( 0,0\right)
}^{\left( N-1,N-1\right) }.  \label{4.3}
\end{equation}%
\emph{\ Then (\ref{4.2}) and (\ref{4.3}) imply that }$\det W_{N}=1.$\emph{\
Therefore, the matrix }$W_{N}$\emph{\ is invertible with its inverse }$%
W_{N}^{-1}$\emph{.}

Note that analogs of the matrix $W_{N}$ for classical orthonormal
polynomials as well as for the basis of trigonometric functions do not have
inverses since the first function in both cases is a constant with its
derivative being identical zero.

\section{Transformation Procedure for Inverse Problem 1}

\label{sec:5}

Prior formulating the desired stability estimates and uniqueness theorem for
the Inverse Problem 1 posed in section 2, we need first to apply a certain
transformation procedure to this problem, which is done in this section.
Recall that the coefficient $a_{\alpha _{0}}\left( x,t\right) $ is unknown
for a certain multiindex $\alpha _{0}.$ Below in this section the function $%
u\left( x,t\right) $ is a solution of equation (\ref{2.9}) with the initial
condition (\ref{2.10}). Since we work below with $D_{x}^{\alpha _{0}}u\left(
x,t\right) $ and since it might happen that $\left\vert \alpha
_{0}\right\vert =2,$ then, keeping in mind Remark 3.1, we assume below that
there exists the derivative $u_{x_{0}}\left( x,t,x_{0}\right) $ and 
\begin{equation}
u\left( x,t,x_{0}\right) ,u_{x_{0}}\left( x,t,x_{0}\right) \in C^{6,3}\left( 
\overline{Q}_{T}\right) \times C\left[ a,b\right] .  \label{5.0}
\end{equation}%
We assume that there exists a number $c>0$ such that 
\begin{equation}
\left\vert D_{x}^{\alpha _{0}}u\left( x,t,x_{0}\right) \right\vert \geq c,%
\text{ }\forall \left( x,t\right) \in \overline{Q}_{T}\times \left[ a,b%
\right] .  \label{5.1}
\end{equation}

To guarantee (\ref{5.1}), some sufficient conditions can be imposed
sometimes.\emph{\ }For example, if $\alpha _{0}=\left( 0,...,0\right) ,$ $%
a_{\alpha _{0}}\leq 0,$ and if the initial boundary value problem of section
3 is considered, then (\ref{5.1}) can be verified using the maximum
principle via imposing corresponding conditions on functions $f$ and $p$ 
\cite[chapter 2]{F}. Another option here is to assume that the number $T$ is
sufficiently small and that 
\begin{equation*}
\left\vert D_{x}^{\alpha _{0}}f\left( x,x_{0}\right) \right\vert \geq 2c,%
\text{ }\forall \left( x,x_{0}\right) \in \overline{\Omega }\times \left[ a,b%
\right] .
\end{equation*}%
Thus, we are not discussing anymore sufficient conditions guaranteeing (\ref%
{5.1}).

Suppose that there exist two solutions of IP1, $\left( a_{\alpha
_{0}}^{\left( 1\right) }\left( x,t\right) ,f_{1}\left( x,x_{0}\right)
\right) $ and

$\left( a_{\alpha _{0}}^{\left( 2\right) }\left( x,t\right) ,f_{2}\left(
x,x_{0}\right) \right) $. $\ $ Let two functions $u_{1}\left(
x,t,x_{0}\right) $ and $u_{2}\left( x,t,x_{0}\right) $ satisfying condition (%
\ref{5.0}), (\ref{5.1}) be two solutions of equation (\ref{2.9}) with
coefficients $a_{\alpha _{0}}^{\left( 1\right) }\left( x,t\right) $ and $%
a_{\alpha _{0}}^{\left( 2\right) }\left( x,t\right) $ respectively as well
as with initial conditions in (\ref{2.10}) $f_{1}\left( x,x_{0}\right) $ and 
$f_{2}\left( x,x_{0}\right) $ respectively. As to the lateral Cauchy data (%
\ref{2.11}), let%
\begin{equation}
u_{k}\mid _{\Gamma _{T}}=g_{0,k}\left( x,t,x_{0}\right) ,\text{ }\partial
_{n}u_{k}\mid _{\Gamma _{T}}=g_{1,k}\left( x,t,x_{0}\right) ,\text{ }%
x_{0}\in \left( a,b\right) ,\text{ }k=1,2.  \label{5.2}
\end{equation}%
Denote 
\begin{equation}
\left. 
\begin{array}{c}
\widetilde{u}\left( x,t,x_{0}\right) =u_{1}\left( x,t,x_{0}\right)
-u_{2}\left( x,t,x_{0}\right) , \\ 
\widetilde{a}_{\alpha _{0}}\left( x,t\right) =a_{\alpha _{0}}^{\left(
1\right) }\left( x,t\right) -a_{\alpha _{0}}^{\left( 2\right) }\left(
x,t\right) , \\ 
\widetilde{f}\left( x,x_{0}\right) =f_{1}\left( x,x_{0}\right) -f_{2}\left(
x,x_{0}\right) , \\ 
\widetilde{g}_{0}\left( x,t,x_{0}\right) =g_{0,1}\left( x,t,x_{0}\right)
-g_{0,2}\left( x,t,x_{0}\right) , \\ 
\widetilde{g}_{1}\left( x,t,x_{0}\right) =g_{1,1}\left( x,t,x_{0}\right)
-g_{1,2}\left( x,t,x_{0}\right) .%
\end{array}%
\right.  \label{5.3}
\end{equation}%
In addition, denote 
\begin{equation}
R\left( x,t,x_{0}\right) =D_{x}^{\left\vert \alpha _{0}\right\vert
}u_{1}\left( x,t,x_{0}\right) .  \label{5.4}
\end{equation}%
It follows from (\ref{2.8}), (\ref{5.0}), (\ref{5.1}) and (\ref{5.4}) that%
\begin{equation}
\widetilde{f}\left( x,x_{0}\right) ,\text{ }\partial _{x_{0}}\widetilde{f}%
\left( x,x_{0}\right) \in C\left( \overline{\Omega }\right) \times C\left[
a,b\right] ,  \label{5.5}
\end{equation}%
\begin{equation}
\widetilde{u}\left( x,t,x_{0}\right) ,\text{ }\widetilde{u}_{x_{0}}\left(
x,t,x_{0}\right) \in C^{6,3}\left( \overline{Q}_{T}\right) \times C\left[ a,b%
\right] ,  \label{5.6}
\end{equation}%
\begin{equation}
R\left( x,t,x_{0}\right) ,\text{ }R_{x_{0}}\left( x,t,x_{0}\right) \in
C^{4,2}\left( \overline{Q}_{T}\right) \times C\left[ a,b\right] ,
\label{5.7}
\end{equation}%
\begin{equation}
\left\vert R\left( x,t,x_{0}\right) \right\vert \geq c,\text{ }\forall
\left( x,t,x_{0}\right) \in \overline{Q}_{T}\times \left[ a,b\right] .
\label{5.8}
\end{equation}%
Using (\ref{2.9})-(\ref{2.11}) and (\ref{5.3}), we obtain%
\begin{equation}
\widetilde{u}_{t}-L\widetilde{u}=R\left( x,t,x_{0}\right) \widetilde{a}%
_{\alpha _{0}}\left( x,t\right) ,\text{ }\left( x,t,x_{0}\right) \in
Q_{T}\times \left( a,b\right) ,  \label{5.9}
\end{equation}%
\begin{equation}
\widetilde{u}\left( x,0,x_{0}\right) =\widetilde{f}\left( x,x_{0}\right)
,x_{0}\in \left( a,b\right) ,  \label{5.10}
\end{equation}%
\begin{equation}
\widetilde{u}\mid _{\Gamma _{T}}=\widetilde{g}_{0}\left( x,t,x_{0}\right) ,%
\text{ }\partial _{n}\widetilde{u}\mid _{\Gamma _{T}}=\widetilde{g}%
_{1}\left( x,t,x_{0}\right) ,\text{ }x_{0}\in \left( a,b\right) .
\label{5.11}
\end{equation}%
Using (\ref{5.8}), divide both sides of equation (\ref{5.9}) by $R\left(
x,t,x_{0}\right) $ and denote%
\begin{equation}
v\left( x,t,x_{0}\right) =\frac{\widetilde{u}\left( x,t,x_{0}\right) }{%
R\left( x,t,x_{0}\right) }.  \label{5.12}
\end{equation}%
Then equation (\ref{5.9}) with the initial condition (\ref{5.10}) and
lateral Cauchy data (\ref{5.11}) become:%
\begin{equation}
v_{t}-L_{0}v-Pv=\widetilde{a}_{\alpha _{0}}\left( x,t\right) ,\text{ }\left(
x,t,x_{0}\right) \in Q_{T}\times \left( a,b\right) ,  \label{5.13}
\end{equation}%
\begin{equation}
v\left( x,0,x_{0}\right) =\frac{\widetilde{f}\left( x,x_{0}\right) }{R\left(
x,0,x_{0}\right) },\text{ }x_{0}\in \left( a,b\right) ,  \label{5.14}
\end{equation}%
\begin{equation}
v\mid _{\Gamma _{T}}=\overline{g}_{0}\left( x,t,x_{0}\right) ,\text{ }%
\partial _{n}v\mid _{\Gamma _{T}}=\overline{g}_{1}\left( x,t,x_{0}\right) ,%
\text{ }x_{0}\in \left( a,b\right) .  \label{5.15}
\end{equation}%
In (\ref{5.15})%
\begin{equation}
\overline{g}_{0}\left( x,t,x_{0}\right) =\frac{\widetilde{g}_{0}}{R}\left(
x,t,x_{0}\right) ,\text{ }\overline{g}_{1}\left( x,t,x_{0}\right) =\left( 
\frac{\widetilde{g}_{1}}{R}-\frac{\partial _{n}R}{R^{2}}\widetilde{g}%
_{0}\right) \left( x,t,x_{0}\right) .  \label{5.16}
\end{equation}%
Hence, it follows from (\ref{5.0})-(\ref{5.3}), (\ref{5.7}), (\ref{5.8}), (%
\ref{5.15}) and (\ref{5.16}) that%
\begin{equation}
\overline{g}_{i}\left( x,t,x_{0}\right) ,\partial _{x_{0}}\overline{g}%
_{i}\left( x,t,x_{0}\right) \in C^{4,2}\left( \overline{\Gamma }_{T}\right)
\times C\left[ a,b\right] ,\text{ }i=1,2.  \label{5.17}
\end{equation}%
The operator $L_{0}$ in (\ref{5.13}) is the same as the one in (\ref{2.3})
with the properties (\ref{2.6}), (\ref{2.5}), and the operator $P$ in (\ref%
{5.13}) is%
\begin{equation}
Pv=\sum\limits_{\left\vert \alpha \right\vert \leq 1}b_{\alpha }\left(
x,t,x_{0}\right) D_{x}^{\alpha }v.  \label{5.18}
\end{equation}%
It follows from (\ref{2.6}), (\ref{5.7}), (\ref{5.8}) and (\ref{5.12}) that
coefficients of the operator $P$ satisfy the following conditions%
\begin{equation}
b_{\alpha }\left( x,t,x_{0}\right) ,\text{ }\partial _{x_{0}}b_{\alpha
}\left( x,t,x_{0}\right) \in C\left( \overline{Q}_{T}\right) \times C\left[
a,b\right] .  \label{5.19}
\end{equation}%
Furthermore, there exists a number 
\begin{equation}
\overline{C}=\overline{C}\left( B,\left\Vert R\right\Vert _{C^{2,1}\left( 
\overline{Q}_{T}\right) },\left\Vert \partial _{x_{0}}R\right\Vert
_{C^{2,1}\left( \overline{Q}_{T}\right) },c\right) >0  \label{5.190}
\end{equation}%
such that 
\begin{equation}
\max_{\left\vert \alpha \right\vert \leq 1}\left\Vert b_{\alpha }\right\Vert
_{C\left( \overline{Q}_{T}\right) \times C\left[ a,b\right] },\text{ }%
\max_{\left\vert \alpha \right\vert \leq 1}\left\Vert \partial
_{x_{0}}b_{\alpha }\right\Vert _{C\left( \overline{Q}_{T}\right) \times C%
\left[ a,b\right] }\leq \overline{C},  \label{5.191}
\end{equation}%
where $B>0$ is the number in (\ref{2.6}). The number $c>0$ in (\ref{5.190})
is the one from (\ref{5.8}). Below $\overline{C}>0$ denotes different
numbers depending on parameters listed in (\ref{5.190}). It follows from (%
\ref{5.6})-(\ref{5.8}) and (\ref{5.12}) that 
\begin{equation}
v\left( x,t,x_{0}\right) ,\text{ }\partial _{x_{0}}v\left( x,t,x_{0}\right)
\in C^{4,2}\left( \overline{Q}_{T}\right) \times C\left[ a,b\right] .
\label{5.180}
\end{equation}

Let $\{\Psi _{k}\left( x_{0}\right) \}_{k=0}^{\infty }$ be the set of
functions (\ref{4.1}), which form the orthonormal basis in $L_{2}\left(
a,b\right) $ described in section 4. Let $N>1$ be an arbitrary integer. We
assume below that the function $v\left( x,t,x_{0}\right) $ has the form%
\begin{equation}
v\left( x,t,x_{0}\right) =\sum\limits_{j=0}^{N-1}v_{j}\left( x,t\right)
\Psi _{j}\left( x_{0}\right) ,\text{ }\left( x,t,x_{0}\right) \in \overline{Q%
}_{T}\times \left( a,b\right) ,  \label{6.1}
\end{equation}%
where coefficients $v_{j}\left( x,t\right) $ of expansion (\ref{6.1}) are
unknown. It follows from (\ref{5.180}) that 
\begin{equation}
v_{j}\in C^{4,2}\left( \overline{Q}_{T}\right) ,\text{ }j=0,...,N-1.
\label{6.2}
\end{equation}%
Furthermore, we assume that the substitution of (\ref{6.1}) in equation (\ref%
{5.13}) still keeps this equation valid, i.e.%
\begin{equation}
\left. \sum\limits_{j=0}^{N-1}\left( \partial _{t}-L_{0}-P\right) \left(
v_{j}\left( x,t\right) \right) \Psi _{j}\left( x_{0}\right) =\widetilde{a}%
_{\alpha _{0}}\left( x,t\right) ,\text{ }\left( x,t,x_{0}\right) \in
Q_{T}\times \left( a,b\right) .\right.  \label{6.3}
\end{equation}%
In addition, we assume that equation (\ref{6.3}) can be differentiated once
with respect to $x_{0}$ and that the resulting equation is still valid, 
\begin{equation}
\left. 
\begin{array}{c}
\sum\limits_{j=0}^{N-1}\left( \partial _{t}-L_{0}-P\right) \left(
v_{j}\left( x,t\right) \right) \Psi _{j}^{\prime }\left( x_{0}\right) - \\ 
-\sum\limits_{\left\vert \alpha \right\vert \leq 1}\partial
_{x_{0}}b_{\alpha }\left( x,t,x_{0}\right) \cdot \left(
\sum\limits_{j=0}^{N-1}D_{x}^{\alpha }v_{j}\left( x,t\right) \cdot \Psi
_{j}\left( x_{0}\right) \right) =0, \\ 
\left( x,t,x_{0}\right) \in Q_{T}\times \left( a,b\right) .%
\end{array}%
\right.  \label{6.4}
\end{equation}%
where we have used (\ref{5.18}) and (\ref{5.19}).

Using (\ref{5.14}) (\ref{6.1}), we obtain for the initial conditions for the
functions $v_{j}\left( x,t\right) $ 
\begin{equation}
\left. 
\begin{array}{c}
\widetilde{f}\left( x,x_{0}\right) /R\left( x,0,x_{0}\right)
=\sum\limits_{j=0}^{N-1}v_{j}\left( x,0\right) \Psi _{j}\left( x_{0}\right)
,\text{ }\left( x,x_{0}\right) \in \overline{\Omega }\times \left(
a,b\right) , \\ 
v_{j}\left( x,0\right) =\int\limits_{a}^{b}\left( \widetilde{f}\left(
x,x_{0}\right) /R\left( x,0,x_{0}\right) \right) \Psi _{j}\left(
x_{0}\right) dx_{0},\text{ }x\in \Omega ,\text{ }j=0,...,N-1.%
\end{array}%
\right.  \label{6.40}
\end{equation}%
In addition, using (\ref{2.1}), (\ref{2.2}), (\ref{5.15})-(\ref{5.17}) and (%
\ref{6.1}), we obtain for the lateral Cauchy data for functions $v_{j}\left(
x,t\right) $ at $\Gamma _{T}:$%
\begin{equation}
\left. 
\begin{array}{c}
\overline{g}_{0}\left( x,t,x_{0}\right) =\sum\limits_{j=0}^{N-1}\overline{g}%
_{0,j}\left( x,t\right) \Psi _{j}\left( x_{0}\right) ,\text{ }\left(
x,t,x_{0}\right) \in \Gamma _{T}\times \left( a,b\right) , \\ 
\overline{g}_{1}\left( x,t,x_{0}\right) =\sum\limits_{j=0}^{N-1}\overline{g}%
_{1,j}\left( x,t\right) \Psi _{j}\left( x_{0}\right) ,\text{ }\left(
x,t,x_{0}\right) \in \Gamma _{T}\times \left( a,b\right) , \\ 
\overline{g}_{0,j}\left( x,t\right) =\int\limits_{a}^{b}\overline{g}%
_{0}\left( x,t,x_{0}\right) \Psi _{j}\left( x_{0}\right) dx_{0},\text{ }%
\left( x,t\right) \in \Gamma _{T}, \\ 
\overline{g}_{1,j}\left( x,t\right) =\int\limits_{a}^{b}\overline{g}%
_{1}\left( x,t,x_{0}\right) \Psi _{j}\left( x_{0}\right) dx_{0},\text{ }%
\left( x,t\right) \in \Gamma _{T}, \\ 
v_{j}\mid _{\Gamma _{T}}=\overline{g}_{0,j}\left( x,t\right) ,\text{ }%
\partial _{n}v_{j}\mid _{\Gamma _{T}}=\overline{g}_{1,j}\left( x,t\right) ,
\\ 
j=0,...,N-1.%
\end{array}%
\right.  \label{6.5}
\end{equation}

The transformation procedure for IP1 is completed.

\section{Formulations of Stability and Uniqueness Results for Inverse
Problem 1}

\label{sec:6}

As to (\ref{2.1}), it is shown in subsection 7.1 that one can assume,
without a loss of generality that 
\begin{equation}
\Omega \subset \left\{ x_{1}\in \left( 0,\frac{1}{8}\right) \right\} .
\label{6.07}
\end{equation}%
Thus, keeping in mind (\ref{6.07}), consider two numbers $\gamma _{1}$ and $%
\gamma _{2}$ such that 
\begin{equation}
\left. 
\begin{array}{c}
\gamma _{2}>\gamma _{1}>0, \\ 
\text{ }\gamma _{2}-\gamma _{1}\in \left( 0,1/8\right) .\text{ }%
\end{array}%
\right.  \label{6.7}
\end{equation}%
By (\ref{6.7}) $\sqrt{2\left( \gamma _{2}-\gamma _{1}\right) }<1-\sqrt{%
2\left( \gamma _{2}-\gamma _{1}\right) }.$ Choose a number $t_{0}$ such that 
\begin{equation}
t_{0}\in \left( T\sqrt{2\left( \gamma _{2}-\gamma _{1}\right) },T\left( 1-%
\sqrt{2\left( \gamma _{2}-\gamma _{1}\right) }\right) \right) \subset \left(
0,T\right) .  \label{6.8}
\end{equation}%
Recalling (\ref{2.1}), (\ref{2.2}) and (\ref{6.07}), consider the function $%
\psi _{1}\left( x,t\right) $ which depends on the numbers $Y,t_{0},T,\gamma
_{1}$ as on parameters,%
\begin{equation}
\psi _{1}\left( x,t\right) =x_{1}+\frac{\left\vert y\right\vert ^{2}}{2Y^{2}}%
+\frac{\left( t-t_{0}\right) ^{2}}{2T^{2}}+\gamma _{1},\text{ }\left(
x,t\right) \in Q_{T}.  \label{6.9}
\end{equation}%
We use this function below in our first Carleman estimate in subsection 8.1.
Using (\ref{6.7})-(\ref{6.9}), consider the subdomain $Q_{T}\left( \gamma
_{1},\gamma _{2},t_{0}\right) $ of the domain $Q_{T},$%
\begin{equation}
Q_{T}\left( \gamma _{1},\gamma _{2},t_{0}\right) =\left\{ \left( x,t\right)
\in Q_{T}:\psi _{1}\left( x,t\right) <\gamma _{2}\right\} .  \label{6.10}
\end{equation}%
It is shown in subsection 7.1 that (\ref{6.10}) does not imply that the size
of the domain $Q_{T}\left( \gamma _{1},\gamma _{2},t_{0}\right) $ is small
in the $x_{1}-$direction. By (\ref{6.7})-(\ref{6.10}) 
\begin{equation}
\left. 
\begin{array}{c}
\overline{Q_{T}\left( \gamma _{1},\gamma _{2},t_{0}\right) }\cap \left\{
t=0\right\} =\varnothing , \\ 
\overline{Q_{T}\left( \gamma _{1},\gamma _{2},t_{0}\right) }\cap \left\{
t=T\right\} =\varnothing .%
\end{array}%
\right.  \label{6.11}
\end{equation}%
In addition, (\ref{6.9})-(\ref{6.11}) imply that the boundary $\partial
Q_{T}\left( \gamma _{1},\gamma _{2},\overline{t}\right) $ of the domain $%
Q_{T}\left( \gamma _{1},\gamma _{2},\overline{t}\right) $ consists of two
parts,%
\begin{equation}
\partial Q_{T}\left( \gamma _{1},\gamma _{2},t_{0}\right) =\partial
_{1}Q_{T}\left( \gamma _{1},\gamma _{2},t_{0}\right) \cup \partial
_{2}Q_{T}\left( \gamma _{1},\gamma _{2},t_{0}\right) ,  \label{6.120}
\end{equation}%
\begin{equation}
\partial _{1}Q_{T}\left( \gamma _{1},\gamma _{2},t_{0}\right) =\left\{
\left( x,t\right) :\psi _{1}\left( x,t\right) =\gamma _{2}\right\} \subset
Q_{T},  \label{6.121}
\end{equation}%
\begin{equation}
\left. 
\begin{array}{c}
\partial _{2}Q_{T}\left( \gamma _{1},\gamma _{2},t_{0}\right) = \\ 
=\left\{ \left( x,t\right) =\left( 0,y,t\right) :\left\vert y\right\vert
^{2}/\left( 2Y\right) ^{2}+\left( t-t_{0}\right) ^{2}/\left( 2T\right)
^{2}<\gamma _{2}-\gamma _{1}\right\} \subset \Gamma _{T}.%
\end{array}%
\right.  \label{6.12}
\end{equation}%
Hence, the domain $Q_{T}\left( \gamma _{1},\gamma _{2},t_{0}\right) $ is a
part of a paraboloid $\partial _{1}Q_{T}\left( \gamma _{1},\gamma
_{2},t_{0}\right) $ is a part of the boundary of that paraboloid and, at the
same time, it is a part of the level surface of the function $\psi
_{1}\left( x,t\right) .$ Consider a sufficiently small number $\omega \in
\left( 0,\gamma _{2}-\gamma _{1}\right) $ and the subdomain $Q_{T}^{\omega
}\left( \gamma _{1},\gamma _{2},t_{0}\right) $ of the domain $Q_{T}\left(
\gamma _{1},\gamma _{2},t_{0}\right) ,$%
\begin{equation}
Q_{T}^{\omega }\left( \gamma _{1},\gamma _{2},t_{0}\right) =\left\{ \left(
x,t\right) \in Q_{T}:\psi _{1}\left( x,t\right) <\gamma _{2}-\omega \right\}
\subset Q_{T}\left( \gamma _{1},\gamma _{2},t_{0}\right) .  \label{6.13}
\end{equation}

By (\ref{6.2}) there exists a number $K>0$ such that 
\begin{equation}
\max_{j\in \left[ 0,N-1\right] }\left\Vert v_{j}\right\Vert _{C^{2,1}\left( 
\overline{Q}_{T}\right) }\leq K.  \label{6.14}
\end{equation}%
Below $H^{k,s}$ are well known Hilbert spaces, where $k,s\geq 0,k+s\geq 1$
are integers. We assume in this section below that conditions (\ref{2.1})-(%
\ref{2.8}) as well as those formulated in section 5 hold.

\textbf{Theorem 6.1} (H\"{o}lder stability estimate for the function $%
\widetilde{a}_{\alpha _{0}}\left( x,t\right) ).$ \emph{Assume that
conditions (\ref{6.1})-(\ref{6.10}), (\ref{6.13}) and (\ref{6.14}) hold.
Suppose that }%
\begin{equation}
\max_{j\in \left[ 0,N-1\right] }\left\Vert \overline{g}_{0,j}\right\Vert
_{H^{1,1}\left( \Gamma _{T}\right) },\text{ }\max_{j\in \left[ 0,N-1\right]
}\left\Vert \overline{g}_{1,j}\right\Vert _{L_{2}\left( \Gamma _{T}\right)
}\leq \delta ,  \label{6.15}
\end{equation}%
\emph{where }$\delta >0$\emph{\ is a sufficiently small number. Let }$%
\overline{C}>0$\emph{\ be the number defined in (\ref{5.190}), (\ref{5.191}%
). Then there exists a number }%
\begin{equation}
C_{1}=C_{1}\left( \mu ,N,\left( a,b\right) ,Q_{T},K,t_{0},\overline{C}%
,\omega ,\gamma _{1},\gamma _{2}\right) >0  \label{6.16}
\end{equation}%
\emph{a number }$\rho _{1}\in \left( 0,1/6\right) $\emph{\ and a
sufficiently small number} $\delta _{1}\in \left( 0,1\right) ,$ \emph{all
three numbers }$C_{1},\rho _{1}$\emph{\ and }$\delta _{1}$\emph{\ depending
only on parameters listed in (\ref{6.16}), such that the following H\"{o}%
lder stability estimates are valid}%
\begin{equation}
\max_{j\in \left[ 0,N-1\right] }\left\Vert v_{j}\right\Vert _{H^{1,0}\left(
Q_{T}^{\omega }\left( \gamma _{1},\gamma _{2},t_{0}\right) \right) }\leq
C_{1}\delta ^{\rho _{1}},\text{ }\forall \delta \in \left( 0,\delta
_{1}\right) ,  \label{6.17}
\end{equation}%
\begin{equation}
\left\Vert \widetilde{a}_{\alpha _{0}}\right\Vert _{L_{2}\left(
Q_{T}^{\omega }\left( \gamma _{1},\gamma _{2},t_{0}\right) \right) }\leq
C_{1}\delta ^{\rho _{1}},\text{ }\forall \delta \in \left( 0,\delta
_{1}\right) .  \label{6.18}
\end{equation}%
\emph{The dependence on }$\overline{C}$\emph{\ in (\ref{6.16}) means the
dependence on the same parameters as ones listed in (\ref{5.190}).}

Below $C_{1}>0$ denotes different numbers depending on parameters listed in (%
\ref{6.16}).

\textbf{Theorem 6.2 }(uniqueness). \emph{Assume that all conditions of
Theorem 6.1 hold, and also that }$\delta =0$ \emph{in (\ref{6.15}). Then}%
\begin{equation}
\left. 
\begin{array}{c}
v_{j}\left( x,t\right) =0\text{ \emph{in} }Q_{T},\text{ }j=0,...,N-1, \\ 
\widetilde{a}_{\alpha _{0}}\left( x,t\right) =0\text{ \emph{in} }Q_{T},%
\end{array}%
\right.  \label{6.21}
\end{equation}%
\begin{equation}
\widetilde{f}\left( x,x_{0}\right) =0\text{ \emph{in} }\left( x,x_{0}\right)
\in \Omega \times \left( a,b\right) .  \label{6.22}
\end{equation}

We now prove Theorem 6.2, assuming that Theorem 6.1 holds true.

\textbf{Proof of Theorem 6.2.} Since $\omega \in \left( 0,\gamma _{2}-\gamma
_{1}\right) $ is an arbitrary small number in Theorem 6.1 and since $\delta
=0$, then (\ref{6.13}), (\ref{6.17}) and (\ref{6.18}) imply 
\begin{equation}
\left. 
\begin{array}{c}
v_{j}\left( x,t\right) =0\text{ in }Q_{T}\left( \gamma _{1},\gamma
_{2},t_{0}\right) ,\text{ }j=0,...,N-1, \\ 
\widetilde{a}_{\alpha _{0}}\left( x,t\right) =0\text{ in }Q_{T}\left( \gamma
_{1},\gamma _{2},t_{0}\right) .%
\end{array}%
\right.  \label{6.23}
\end{equation}%
It obviously follows from (\ref{6.7})-(\ref{6.10}) that rotating the
coordinate axis and moving the coordinate system, one can cover the entire
open domain $Q_{T}$ by its subdomains like $Q_{T}\left( \gamma _{1},\gamma
_{2},t_{0}\right) $ in (\ref{6.10}), and all these subdomains will have
property (\ref{6.11}). Thus, (\ref{6.23}) implies that, doing this
sequentially in an obvious manner, we obtain (\ref{6.21}). Setting $t=0$ in
the first line of (\ref{6.21}) and using (\ref{6.40}), we obtain (\ref{6.22}%
). $\square $

\section{Stability Estimates for Both the Unknown Coefficient and the
Initial Condition}

\label{sec:7}

In the case of IP1 a H\"{o}lder stability estimate was formulated in Theorem
6.1 for the unknown coefficient. On the other hand, only uniqueness result
for the initial condition was formulated in Theorem 6.2. Unlike this, we
consider in the current section Inverse Problem 2 (IP2), for which we derive
stability estimates for both: the unknown coefficient and the initial
condition.

\subsection{Statement of Inverse Problem 2}

\label{sec:7.1}

Based on (\ref{2.1}), we assume in this section that the domain $\Omega $ is
a rectangular prism and that the hypersurface $\Gamma $ is one of its sides,%
\begin{equation}
\left. 
\begin{array}{c}
\Omega =\left\{ x=\left( x_{1},y\right) :x_{1}\in \left( 0,A\right) ,\text{ }%
\left\vert y_{i}\right\vert <Y,\text{ }i=1,...,n-1\right\} , \\ 
\Gamma =\left\{ x=\left( x_{1},y\right) :x_{1}=0,\text{ }\left\vert
y_{i}\right\vert <Y,\text{ }i=1,...,n-1\right\} \subset \partial \Omega ,%
\end{array}%
\right.  \label{6.130}
\end{equation}%
where the number $A>0.$ We assume below that in (\ref{6.130}) 
\begin{equation}
A\in \left( 0,\frac{1}{8}\right) .  \label{7.1}
\end{equation}%
The generality is not lost due to assumption (\ref{7.1}). Indeed, we can
change variables in equation (\ref{2.9}) as 
\begin{equation*}
x_{1}^{\prime }=\frac{s}{A}x_{1},\text{ }y^{\prime }=\frac{s}{A}y,\text{ }%
t^{\prime }=\frac{s^{2}}{A^{2}}y,
\end{equation*}%
where $s\in \left( 0,1/8\right) $ is any number. Then $x_{1}^{\prime }\in
\left( 0,s\right) .$ Next, dividing both sides of the resulting equation by $%
\left( s^{2}/A^{2}\right) ,$ we obtain a new parabolic equation, in which,
however, the principal part $\left( \partial _{t^{\prime }}-L_{0}\right) $
of the parabolic operator remains the same as the one in (\ref{2.3}), (\ref%
{2.30}), (\ref{2.9}). Since Carleman estimates are concerned only with
principal parts of PDE operators \cite[Lemma 2.1.1]{KL}, \cite[chapter 4]%
{LRS}, then the unchanged principal part will not affect our derivations
below.

Inverse Problem 2 is a specification of Inverse Problem 1 for the case of
the domain $\Omega $ in (\ref{6.130}) and under the assumption that the
Neumann boundary condition is given at $\partial \Omega \times \left(
d_{2},T\right) ,$ where $d_{2}\in \left( 0,T\right) $ is a certain number
specified below in this subsection 7.1$.$ Also, keeping in mind the
transformation procedure of section 5, we replace now regularity requirement
(\ref{2.80}) with the regularity requirement (\ref{5.0}).

Using (\ref{7.1}), consider an arbitrary number $\gamma >0$ such that 
\begin{equation}
\gamma +A\in \left( 0,\frac{1}{8}\right) .  \label{7.100}
\end{equation}%
In particular, (\ref{7.1}) and (\ref{7.100}) imply%
\begin{equation}
\left. 
\begin{array}{c}
T\left( 1/2+\sqrt{2\gamma }\right) \in \left( 0,T\right) , \\ 
\text{ }T\left( 1/2-\sqrt{2\gamma }\right) =d_{1}>0, \\ 
T\left( 1/2-\sqrt{2\left( \gamma +A\right) }\right) =d_{2}\in \left(
0,d_{1}\right) .%
\end{array}%
\right.  \label{7.101}
\end{equation}%
Using (\ref{2.2}) and (\ref{7.101}), denote 
\begin{equation}
S_{T,d_{2}}=\partial \Omega \times \left( d_{2},T\right) .  \label{7.102}
\end{equation}%
Let 
\begin{equation}
\varepsilon \in \left( 0,\frac{\gamma }{2}\right)  \label{7.1020}
\end{equation}%
be an arbitrary number. Denote 
\begin{equation}
\left. 
\begin{array}{c}
t_{\varepsilon }^{+}=T\left( 1/2+\sqrt{2\left( \gamma -2\varepsilon \right) }%
\right) , \\ 
t_{\varepsilon }^{-}=T\left( 1/2-\sqrt{2\left( \gamma -2\varepsilon \right) }%
\right) >d_{1}, \\ 
P\left( t_{\varepsilon }^{-},T\right) =\Omega \times \left( t_{\varepsilon
}^{-},T\right) .%
\end{array}%
\right.  \label{7.103}
\end{equation}%
By (\ref{7.101}), (\ref{7.1020}) and (\ref{7.103})%
\begin{equation}
\left. 
\begin{array}{c}
0<t_{\varepsilon }^{-}<t_{\varepsilon }^{+}<T, \\ 
t_{\varepsilon }^{+}-t_{\varepsilon }^{-}=2\sqrt{2\left( \gamma
-2\varepsilon \right) }=d_{\varepsilon }.%
\end{array}%
\right.  \label{7.104}
\end{equation}

\textbf{Inverse Problem 2 (IP2).} \emph{Let conditions (\ref{2.8}), (\ref%
{6.130})-(\ref{7.103}) hold. Let }$\alpha _{0},$ $\left\vert \alpha
_{0}\right\vert \leq 2$\ \emph{be a fixed multiindex. Let the function }$%
u\left( x,t,x_{0}\right) $\emph{\ satisfies conditions (\ref{2.9}), (\ref%
{2.10}), (\ref{5.0}) and (\ref{5.1}). Suppose that the coefficient }$%
a_{\alpha _{0}}\left( x,t\right) $\emph{\ of the operator }$L$\emph{\ in (%
\ref{2.3}) is unknown for }$\left( x,t\right) \in P\left( t_{\varepsilon
}^{-},T\right) $\emph{\ and known for }$\left( x,t\right) \in \Omega \times
\left( 0,t_{\varepsilon }^{-}\right) .$\emph{\ In addition, assume that
conditions (\ref{2.11}) are replaced with }%
\begin{equation}
\left. u\mid _{S_{T}}=p_{0}\left( x,t,x_{0}\right) ,\text{ }\partial
_{n}u\mid _{S_{T,d_{2}}}=p_{1}\left( x,t,x_{0}\right) ,\text{ }x_{0}\in
\left( a,b\right) ,\right.  \label{7.2}
\end{equation}%
\emph{where functions }$p_{0}$\emph{\ and }$p_{1}$ \emph{are known and }$%
S_{T,d_{2}}$ \emph{is defined in (\ref{7.102}). Determine the coefficient }$%
a_{\alpha _{0}}\left( x,t\right) $ \emph{for} $\left( x,t\right) \in P\left(
t_{\varepsilon }^{-},T\right) $ \emph{as well as the initial condition }$%
f\left( x,x_{0}\right) $\emph{\ for }$\left( x,x_{0}\right) $\emph{\ }$\in
\Omega \times \left( a,b\right) .$

Thus, while the Neumann boundary condition in (\ref{7.2}) is given at the
entire lateral boundary $S_{T}$ and the Dirichlet boundary condition is
given only on its part $S_{T,d_{2}}\subset S_{T},$ see (\ref{7.101}) and (%
\ref{7.102}). Since in IP2 the coefficient $a_{\alpha _{0}}\left( x,t\right) 
$ is known for $\left( x,t\right) \in \Omega \times \left( 0,t_{\varepsilon
}^{-}\right) ,$ then uniqueness of the solution of IP2 follows immediately
from Theorem 6.2, and, therefore, it is not discussed below.

\textbf{Remark 7.1.} \emph{Of course, the existence of the function }$%
u\left( x,t,x_{0}\right) $\emph{\ satisfying conditions (\ref{2.9}), (\ref%
{2.10}), (\ref{5.0}) and the Dirichlet boundary condition at }$\partial
\Omega \times \left( 0,T\right) $ \emph{can be guaranteed only if the
boundary of the domain }$\Omega $\emph{\ is sufficiently smooth and certain
compatibility conditions hold, see Theorem 3.1 and \cite[Theorem 5.2 of
chapter 4]{Lad}. However, this is not the case of the domain in (\ref{6.130}%
). Hence, we can assume that the forward problem is posed as in section 3,
i.e. in a larger bounded domain }$\Omega ^{\prime }$\emph{\ with a
sufficiently smooth boundary and such that the domain }$\Omega $\emph{\ in (%
\ref{6.130}) is a subdomain of }$\Omega ^{\prime }.$\emph{\ Next, we can
assume that all coefficients of the operator }$L$\emph{\ are known in }$%
\left( \Omega ^{\prime }\diagdown \Omega \right) \times \left( 0,T\right) $%
\emph{, the initial condition }$f\left( x,x_{0}\right) $\emph{\ is also
known in }$\left( \Omega ^{\prime }\diagdown \Omega \right) \times \left(
a,b\right) $\emph{\ and that the input data (\ref{7.2}) are measured. Then
we obtain IP2.}

\subsection{Transformation procedure for Inverse Problem 2}

\label{sec:7.2}

We now briefly modify the transformation procedure of section 5 for IP1 for
the case of IP2. Assume the existence of two solutions of IP2, $\left(
a_{\alpha _{0}}^{\left( 1\right) }\left( x,t\right) ,f_{1}\left(
x,x_{0}\right) \right) $ and

$\left( a_{\alpha _{0}}^{\left( 2\right) }\left( x,t\right) ,f_{2}\left(
x,x_{0}\right) \right) $. Consider again two functions $u_{1}\left(
x,t,x_{0}\right) $ and $u_{2}\left( x,t,x_{0}\right) ,$ which satisfy
conditions of IP2. In particular, they satisfy equation (\ref{2.9}) with
coefficients $a_{\alpha _{0}}^{\left( 1\right) }$ and $a_{\alpha
_{0}}^{\left( 1\right) }$ respectively and with initial conditions $%
f_{1}\left( x,x_{0}\right) $ and $f_{2}\left( x,x_{0}\right) $ in (\ref{2.10}%
) respectively. Using (\ref{7.2}), we replace input data (\ref{5.2}) with%
\begin{equation}
\left. 
\begin{array}{c}
u_{k}\mid _{S_{T}}=p_{0,k}\left( x,t,x_{0}\right) ,\text{ }\left(
x,t,x_{0}\right) \in S_{T}\times \left( a,b\right) , \\ 
\partial _{n}u_{k}\mid _{S_{T,d_{2}}}=p_{1,k}\left( x,t,x_{0}\right) ,\text{ 
}\left( x,t,x_{0}\right) \in S_{T,d_{2}}\times \left( a,b\right) ,\text{ }%
k=1,2.\text{ }%
\end{array}%
\right.  \label{7.3}
\end{equation}%
In the case of IP2 formulas (\ref{5.3}) are replaced with%
\begin{equation}
\left. 
\begin{array}{c}
\widetilde{u}\left( x,t,x_{0}\right) =u_{1}\left( x,t,x_{0}\right)
-u_{2}\left( x,t,x_{0}\right) , \\ 
\widetilde{a}_{\alpha _{0}}\left( x,t\right) =a_{\alpha _{0}}^{\left(
1\right) }\left( x,t\right) -a_{\alpha _{0}}^{\left( 2\right) }\left(
x,t\right) , \\ 
\widetilde{f}\left( x,x_{0}\right) =f_{1}\left( x,x_{0}\right) -f_{2}\left(
x,x_{0}\right) , \\ 
\widetilde{p}_{0}\left( x,t,x_{0}\right) =p_{0,1}\left( x,t,x_{0}\right)
-p_{0,2}\left( x,t,x_{0}\right) , \\ 
\widetilde{p}_{1}\left( x,t,x_{0}\right) =p_{1,1}\left( x,t,x_{0}\right)
-p_{1,2}\left( x,t,x_{0}\right) .%
\end{array}%
\right.  \label{7.03}
\end{equation}%
Formulas (\ref{5.4})-(\ref{5.10}) hold again for IP2. As to (\ref{5.11}), we
use (\ref{7.3}) and (\ref{7.03}) to replace it with 
\begin{equation}
\widetilde{u}\mid _{S_{T}}=\widetilde{p}_{0}\left( x,t,x_{0}\right) ,\text{ }%
\partial _{n}\widetilde{u}\mid _{S_{T,d_{2}}}=\widetilde{p}_{1}\left(
x,t,x_{0}\right) ,\text{ }x_{0}\in \left( a,b\right) .  \label{7.4}
\end{equation}%
Formulas (\ref{5.18})-(\ref{5.191}) remain the same for IP2. As to formulas (%
\ref{5.12})-(\ref{5.17}), we rewrite them now for the convenience of the
reader as:%
\begin{equation}
v\left( x,t,x_{0}\right) =\frac{\widetilde{u}\left( x,t,x_{0}\right) }{%
R\left( x,t,x_{0}\right) },  \label{7.400}
\end{equation}%
\begin{equation}
v_{t}-L_{0}v-Pv=\widetilde{a}_{\alpha _{0}}\left( x,t\right) ,\text{ }\left(
x,t,x_{0}\right) \in Q_{T}\times \left( a,b\right) ,  \label{7.40}
\end{equation}%
\begin{equation}
v\left( x,0,x_{0}\right) =\frac{\widetilde{f}\left( x,x_{0}\right) }{R\left(
x,0,x_{0}\right) },  \label{7.41}
\end{equation}%
\begin{equation}
v\mid _{S_{T}}=\overline{p}_{0}\left( x,t,x_{0}\right) ,\text{ }\partial
_{n}v\mid _{S_{T,d_{2}}}=\overline{p}_{1}\left( x,t,x_{0}\right) ,\text{ }%
x_{0}\in \left( a,b\right) ,  \label{7.42}
\end{equation}%
\begin{equation}
\overline{p}_{0}\left( x,t,x_{0}\right) =\frac{\widetilde{p}_{0}}{R}\left(
x,t,x_{0}\right) ,\text{ }\overline{p}_{1}\left( x,t,x_{0}\right) =\left( 
\frac{\widetilde{p}_{1}}{R}-\frac{\partial _{n}R}{R^{2}}\widetilde{p}%
_{0}\right) \left( x,t,x_{0}\right) ,  \label{7.43}
\end{equation}%
\begin{equation}
\left. 
\begin{array}{c}
\overline{p}_{0}\left( x,t,x_{0}\right) ,\partial _{x_{0}}\overline{p}%
_{0}\left( x,t,x_{0}\right) \in C^{6,3}\left( \overline{S}_{T}\right) \times
C\left[ a,b\right] , \\ 
\overline{p}_{1}\left( x,t,x_{0}\right) ,\partial _{x_{0}}\overline{p}%
_{1}\left( x,t,x_{0}\right) \in C^{4,2}\left( \overline{S}_{T,d_{2}}\right)
\times C\left[ a,b\right] .%
\end{array}%
\right.  \label{7.44}
\end{equation}

We now again assume expansion (\ref{6.1}). Formula (\ref{5.180}) again
implies the validity of (\ref{6.2}). Finally, we again assume the validity
of (\ref{6.3})-(\ref{6.40}) and (\ref{6.14}). Using (\ref{7.42}) and (\ref%
{7.43}), we replace formulas (\ref{6.5}) with:%
\begin{equation}
\left. 
\begin{array}{c}
\overline{p}_{0}\left( x,t,x_{0}\right) =\sum\limits_{i=0}^{N-1}\overline{p}%
_{0,j}\left( x,t\right) \Psi _{j}\left( x_{0}\right) ,\text{ }x_{0}\in
\left( a,b\right) , \\ 
\overline{p}_{1}\left( x,t,x_{0}\right) =\sum\limits_{i=0}^{N-1}\overline{p}%
_{1,j}\left( x,t\right) \Psi _{j}\left( x_{0}\right) ,\text{ }x_{0}\in
\left( a,b\right) , \\ 
\overline{p}_{0,j}\left( x,t\right) =\int\limits_{a}^{b}\overline{p}%
_{0}\left( x,t,x_{0}\right) \Psi _{j}\left( x_{0}\right) dx_{0}, \\ 
\overline{p}_{1,j}\left( x,t\right) =\int\limits_{a}^{b}\overline{p}%
_{1}\left( x,t,x_{0}\right) \Psi _{j}\left( x_{0}\right) dx_{0}, \\ 
v_{j}\mid _{S_{T}}=\overline{p}_{0,j}\left( x,t\right) ,\text{ }\partial
_{n}v_{j}\mid _{S_{T,d_{2}}}=\overline{p}_{1,j}\left( x,t\right) , \\ 
j=0,...,N-1.%
\end{array}%
\right.  \label{7.5}
\end{equation}%
Formula (\ref{6.15}) is now replaced with%
\begin{equation}
\max_{j\in \left[ 0,N-1\right] }\left\Vert \overline{p}_{0,j}\right\Vert
_{H^{1,1}\left( S_{T}\right) },\text{ }\max_{j\in \left[ 0,N-1\right]
}\left\Vert \overline{p}_{1,j}\right\Vert _{L_{2}\left( S_{T,d_{2}}\right)
}\leq \delta .  \label{7.8}
\end{equation}

The transformation procedure for Inverse Problem 2 is completed.

\subsection{Formulations of stability estimates for Inverse Problem 2}

\label{sec:7.3}

Similarly with (\ref{6.9}) consider the function $\psi _{2}\left( x,t\right)
,$ which we will use in our second Carleman estimate in subsection 8.2. This
function is 
\begin{equation}
\psi _{2}\left( x,t\right) =x_{1}+\frac{\left( t-T/2\right) ^{2}}{2T^{2}}%
+\gamma ,\text{ }\left( x,t\right) \in Q_{T}.  \label{7.80}
\end{equation}%
Consider the subdomain $G\left( \gamma ,T\right) $ of the domain $Q_{T},$%
\begin{equation}
G\left( \gamma ,T\right) =\left\{ \left( x,t\right) \in Q_{T}:\psi
_{2}\left( x,t\right) <2\gamma +A\right\} .  \label{7.81}
\end{equation}%
It follows from (\ref{7.1})-(\ref{7.101}) that%
\begin{equation}
G\left( \gamma ,T\right) \subset \left( \Omega \times \left( d_{2},T\right)
\right) ,  \label{7.810}
\end{equation}%
\begin{equation}
\overline{G\left( \gamma ,T\right) }\cap \left\{ t=T\right\} =\varnothing ,%
\text{ }\overline{G\left( \gamma ,T\right) }\cap \left\{ t=0\right\}
=\varnothing .  \label{7.82}
\end{equation}%
It follows from (\ref{6.130}), (\ref{7.101}) and (\ref{7.80})-(\ref{7.82})
that the boundary $\partial G\left( \gamma ,T\right) $ of the domain $%
G\left( \gamma ,T\right) $ consists of four parts, 
\begin{equation}
\left. 
\begin{array}{c}
\partial G\left( \gamma ,T\right) =\cup _{k=1}^{4}\partial _{k}G\left(
\gamma ,T\right) , \\ 
\partial _{1}G\left( \gamma ,T\right) =\left\{ 
\begin{array}{c}
\left( x,t\right) :\left( t-T/2\right) ^{2}/\left( 2T^{2}\right) +\gamma
<2\gamma +A, \\ 
\left( x,t\right) \in \Gamma \times \left( d_{2},T\right)%
\end{array}%
\right\} , \\ 
\partial _{2}G\left( \gamma ,T\right) =\left\{ 
\begin{array}{c}
\left( x,t\right) :x_{1}=A,\left\vert y_{i}\right\vert <Y,\text{ }%
i=1,...,n-1, \\ 
\left( t-T/2\right) ^{2}/\left( 2T^{2}\right) +\gamma <2\gamma +A%
\end{array}%
\right\} , \\ 
\partial _{3}G\left( \gamma ,T\right) =\left\{ 
\begin{array}{c}
\left( x,t\right) :\left\vert y_{i}\right\vert =Y,\text{ }i=1,...,n-1, \\ 
x_{1}+\left( t-T/2\right) ^{2}/\left( 2T^{2}\right) +\gamma <2\gamma +A%
\end{array}%
\right\} , \\ 
\partial _{4}G\left( \gamma ,T\right) =\left\{ \left( x,t\right) \in
Q_{T}:\psi _{2}\left( x,t\right) =2\gamma +A\right\} .%
\end{array}%
\right.  \label{7.83}
\end{equation}%
In addition, due to (\ref{7.1020}), (\ref{7.103}) and (\ref{7.81}) denote 
\begin{equation}
G^{\varepsilon }\left( \gamma ,T\right) =\left\{ \left( x,t\right) \in
Q_{T}:\psi _{2}\left( x,t\right) <2\left( \gamma -\varepsilon \right)
+A\right\} \subset G\left( \gamma ,T\right) .  \label{7.84}
\end{equation}

We assume below in this section that conditions formulated in subsections
7.1 and 7.2 as well as conditions (\ref{7.80}), (\ref{7.81}) and (\ref{7.84}%
) hold.

\textbf{Theorem 7.1.} \emph{There exists a number \ }%
\begin{equation}
C_{2}=C_{2}\left( \Omega ,T,\varepsilon ,\gamma ,A,\mu ,N,\left( a,b\right)
,K,\overline{C}\right) >0  \label{7.9}
\end{equation}%
\emph{depending only on listed parameters, a sufficiently small number }$%
\delta _{2}\in \left( 0,1\right) $\emph{\ and a number }$\rho _{2}\in \left(
0,1/6\right) $\emph{, both numbers }$\delta _{2}$\emph{\ and }$\rho _{2}$%
\emph{\ depend on the same parameters ones listed in (\ref{7.9}), such that
the following H\"{o}lder stability estimates hold}%
\begin{equation}
\max_{j\in \left[ 0,N-1\right] }\left\Vert v_{j}\right\Vert _{H^{1,0}\left(
G^{\varepsilon }\left( \gamma ,T\right) \right) }\leq C_{2}\delta ^{\rho
_{2}},\text{ }\forall \delta \in \left( 0,\delta _{2}\right) ,  \label{7.10}
\end{equation}%
\begin{equation}
\left\Vert \widetilde{a}_{\alpha _{0}}\right\Vert _{L_{2}\left(
G^{\varepsilon }\left( \gamma ,T\right) \right) }\leq C_{2}\delta ^{\rho
_{2}},\text{ }\forall \delta \in \left( 0,\delta _{2}\right) .  \label{7.11}
\end{equation}%
\emph{Let }$\sigma \in \left( t_{\varepsilon }^{-},t_{\varepsilon
}^{+}\right) $\emph{\ be an arbitrary number, where the numbers }$%
t_{\varepsilon }^{-},t_{\varepsilon }^{+}$\emph{\ were defined in (\ref%
{7.103}). Assume that }%
\begin{equation}
\widetilde{p}_{0}\left( x,t,x_{0}\right) =0\text{ \emph{for} }\left(
x,t\right) \in \Omega \times \left( 0,\sigma \right) ,\text{ }x_{0}\in
\left( a,b\right) ,  \label{7.110}
\end{equation}%
\emph{where }$\widetilde{p}_{0}\left( x,t,x_{0}\right) $\emph{\ is the
function, which is defined in (\ref{7.4}).} \emph{Let }%
\begin{equation}
\left\Vert \left\vert \nabla f\left( x,x_{0}\right) \right\vert \right\Vert
_{L_{2}\left( \Omega \right) \times C\left[ a,b\right] }\leq F,
\label{7.001}
\end{equation}%
\emph{where }$F>0$\emph{\ is a known number.}

\emph{Then there exists a number }%
\begin{equation}
C_{3}=C_{3}\left( \Omega ,T,\varepsilon ,\gamma ,A,\mu ,N,\left( a,b\right)
,K,\sigma ,\overline{C},F\right) >0  \label{7.01}
\end{equation}%
\emph{and} \emph{a sufficiently small number }$\delta _{3}\in \left(
0,1\right) ,$ \emph{both }$C_{3}$\emph{\ and }$\delta _{3}$\emph{\ depend on
the same parameters as ones listed in (\ref{7.01}), such that the following
logarithmic stability estimate for the initial condition }$\widetilde{f}%
\left( x,x_{0}\right) $ \emph{holds\ }%
\begin{equation}
\left\Vert \widetilde{f}\left( x,x_{0}\right) \right\Vert _{L_{2}\left(
\Omega \right) \times C\left[ a,b\right] }\leq \frac{C_{3}}{\sqrt{\ln \left[
\ln \left( 1/\delta \right) \right] }},\text{ }\forall \emph{\ }\delta \in
\left( 0,\delta _{3}\right) .\text{ }  \label{7.14}
\end{equation}%
\emph{The dependence on }$\overline{C}$\emph{\ in (\ref{7.01}) means the
dependence on the same parameters as ones listed in (\ref{5.190}).}

Below $C_{2}>0$ and $C_{3}>0$ denote different numbers depending on the same
parameters as ones listed in (\ref{7.9}) and (\ref{7.01}) respectively. We
now formulate an extension of Theorem 7.1 for the case when the domain $%
G^{\varepsilon }\left( \gamma ,T\right) $ in (\ref{7.10}), (\ref{7.11}) is
replaced with a larger domain%
\begin{equation}
D\left( \varepsilon ,\gamma ,T\right) =G^{\varepsilon }\left( \gamma
,T\right) \cup \left\{ \Omega \times \left( t_{\varepsilon }^{-},T\right)
\right\} ,  \label{7.15}
\end{equation}%
where the number $t_{\varepsilon }^{-}$ is defined in (\ref{7.103}).

\textbf{Theorem 7.2.}\emph{\ Let conditions of Theorem 7.1 hold. Then
estimates (\ref{7.10}) and (\ref{7.11}) can be replaced with the more
general ones, }%
\begin{equation}
\max_{j\in \left[ 0,N-1\right] }\left\Vert v_{j}\right\Vert _{H^{1,0}\left(
D\left( \varepsilon ,\gamma ,T\right) \right) }\leq C_{2}\delta ^{\rho _{2}},%
\text{ }\forall \delta \in \left( 0,\delta _{2}\right) ,  \label{7.16}
\end{equation}%
\begin{equation}
\left\Vert \widetilde{a}_{\alpha _{0}}\right\Vert _{L_{2}\left( D\left(
\varepsilon ,\gamma ,T\right) \right) }\leq C_{2}\delta ^{\rho _{2}},\text{ }%
\forall \delta \in \left( 0,\delta _{2}\right) ,  \label{7.17}
\end{equation}%
\emph{where the domain }$D\left( \varepsilon ,\gamma ,T\right) $\emph{\ is
defined in (\ref{7.15}) and the numbers }$\delta _{2}\in \left( 0,1\right) $ 
\emph{and }$\rho _{2}\in \left( 0,1\right) $\emph{\ are the same as the ones
is in Theorem 7.1. Estimate (\ref{7.14}) remains valid.}

\textbf{Remark 7.2.} \emph{It follows from (\ref{7.103}) and (\ref{7.15})
that }$P\left( t_{\varepsilon }^{-},T\right) \subset D\left( \varepsilon
,\gamma ,T\right) .$\emph{\ Therefore, Theorem 7.2 addresses IP2 within the
framework described in subsection 7.2..}

\section{Three Carleman Estimates}

\label{sec:8}

To prove Theorems 6.1 and 7.1, we need to use three Carleman estimates for
the parabolic operator $\partial _{t}-L_{0}$. These estimates are described
in the current section. While the first Carleman estimate of Theorems 8.1 is
known, the second and third Carleman estimates of Theorems 8.2 and 8.3
respectively are new. Therefore, we prove Theorems 8.2 and 8.3 here. As
usual, proofs of pointwise Carleman estimates below are space consuming.

\subsection{The first Carleman estimate}

\label{sec:8.1}

Let $\lambda \geq 1$ and $\nu \geq 1$ be two large parameters, which we will
define later. Let $\psi _{1}\left( x,t\right) $ be the function defined in (%
\ref{6.9}). Define the first Carleman Weight Function as:%
\begin{equation}
\varphi _{1,\lambda ,\nu }\left( x,t\right) =\exp \left( 2\lambda \psi
_{1}^{-\nu }\left( x,t\right) \right) ,\text{ }\left( x,t\right) \in
Q_{T}\left( \gamma _{1},\gamma _{2},t_{0}\right) .  \label{8.1}
\end{equation}%
Theorem 8.1 follows immediately from \cite[Theorem 2.3.1]{KL}, \cite[\S 1 of
chapter 4]{LRS}.

\textbf{Theorem 8.1} (the first Carleman estimate). \ \emph{Assume that
conditions (\ref{2.1})- (\ref{2.3}), the ones in the first line of (\ref{2.6}%
), (\ref{2.5}), (\ref{6.7})-(\ref{6.10}) and (\ref{8.1}) hold. Then there
exist sufficiently large numbers }$\nu _{0}$ \emph{and }$\lambda _{0},$ 
\begin{equation}
\nu _{0}=\nu _{0}\left( \mu ,B,Q_{T}\left( \gamma _{1},\gamma
_{2},t_{0}\right) \right) >1,\text{ }\lambda _{0}=\lambda _{0}\left( \mu
,B,Q_{T}\left( \gamma _{1},\gamma _{2},t_{0}\right) \right) >1  \label{8.100}
\end{equation}%
\emph{\ depending only on listed parameters such that the following
pointwise Carleman estimate is valid for all }$\left( x,t\right) \in 
\overline{Q_{T}\left( \gamma _{1},\gamma _{2},t_{0}\right) }$\emph{\ }%
\begin{equation}
\left. 
\begin{array}{c}
\left( u_{t}-L_{0}u\right) ^{2}\varphi _{1,\lambda ,\nu }\geq C_{4}\lambda
\nu \left\vert \nabla u\right\vert ^{2}\varphi _{1,\lambda ,\nu
}+C_{4}\lambda ^{3}\nu ^{4}\psi _{1}^{-2\nu -2}u^{2}\varphi _{1,\lambda ,\nu
}+ \\ 
+\operatorname{div}U+W_{t},\text{ } \\ 
\left\vert U\right\vert +\left\vert W\right\vert \leq C_{4}\lambda ^{3}\nu
^{3}\psi _{1}^{-2\nu -2}\left( \left\vert \nabla u\right\vert
^{2}+u_{t}^{2}+u^{2}\right) \varphi _{1,\lambda ,\nu }, \\ 
\forall \nu \geq \nu _{0},\text{ }\forall \lambda \geq \lambda _{0},\text{ }%
\forall u\in C^{2,1}\left( \overline{Q_{T}\left( \gamma _{1},\gamma
_{2},t_{0}\right) }\right) ,%
\end{array}%
\right.  \label{8.2}
\end{equation}%
\emph{where the number }$C_{4}=C_{4}\left( \mu ,B,Q_{T}\left( \gamma
_{1},\gamma _{2},t_{0}\right) \right) >0$\emph{\ depends only on listed} 
\emph{parameters.}

Below $C_{4}>0$ denotes different numbers depending on parameters listed in
the formulation of this theorem.

\subsection{The second Carleman estimate}

\label{sec:8.2}

Assume that conditions (\ref{6.130})-(\ref{7.104}), (\ref{7.80}) and (\ref%
{7.81}) hold. Again, let $\lambda \geq 1$ and $\nu \geq 1$ be two large
parameters, which we will define later. Define the second Carleman Weight
Function, 
\begin{equation}
\varphi _{2,\lambda ,\nu }\left( x,t\right) =\exp \left( 2\lambda \psi
_{2}^{-\nu }\left( x,t\right) \right) ,\text{ }\left( x,t\right) \in G\left(
\gamma ,T\right) .  \label{8.3}
\end{equation}

\textbf{Theorem 8.2} (the second Carleman estimate). \emph{Assume that
conditions formulated above in this subsection as well as conditions (\ref%
{2.30}), (\ref{2.6}), (\ref{2.5}) and (\ref{8.3}) hold. Then there exist
sufficiently large numbers }%
\begin{equation}
\nu _{1}=\nu _{1}\left( \mu ,B,G\left( \gamma ,T\right) \right) >1\text{ 
\emph{and} }\lambda _{1}=\lambda _{1}\left( \mu ,B,G\left( \gamma ,T\right)
\right) >1  \label{8.04}
\end{equation}%
\emph{depending only on listed parameters such that the following pointwise
Carleman estimate is valid}%
\begin{equation}
\left. 
\begin{array}{c}
\left( u_{t}-L_{0}u\right) ^{2}\varphi _{2,\lambda ,\nu }\geq C_{5}\lambda
\nu \left\vert \nabla u\right\vert ^{2}\varphi _{2,\lambda ,\nu
}+C_{5}\lambda ^{3}\nu ^{4}\psi _{2}^{-2\nu -2}u^{2}\varphi _{2,\lambda ,\nu
}+ \\ 
+\operatorname{div}\widetilde{U}+\widetilde{V}_{t},\text{ } \\ 
\left\vert \widetilde{U}\right\vert +\left\vert \widetilde{V}\right\vert
\leq C_{5}\lambda ^{3}\nu ^{3}\psi _{2}^{-2\nu -2}\left( \left\vert \nabla
u\right\vert ^{2}+u_{t}^{2}+u^{2}\right) \varphi _{2,\lambda ,\nu }, \\ 
\forall \nu \geq \nu _{1},\text{ }\forall \lambda \geq \lambda _{1},\text{ }%
\forall u\in C^{2,1}\left( \overline{G\left( \gamma ,T\right) }\right) ,%
\text{ }\forall \left( x,t\right) \in \overline{G\left( \gamma ,T\right) }.%
\end{array}%
\right.  \label{8.4}
\end{equation}%
where the number $C_{5}=C_{5}\left( \mu ,B,G\left( \gamma ,T\right) \right)
>0$ depends only on listed parameters. \emph{\ }

\textbf{Proof}. We prove this theorem in eight steps. Below $C_{5}$ denotes
different numbers depending on the same parameters as ones listed above.
Temporary change the above notation for $x\in \mathbb{R}^{n}$ as $x=\left(
x_{1},...,x_{n}\right) .$ Below $O\left( 1/\nu \right) $ and $O\left(
1/\lambda \right) $ denote different $C^{1}\left( \overline{G\left( \gamma
,T\right) }\right) -$functions satisfying 
\begin{equation}
\left\vert D^{\alpha }O\left( \frac{1}{\nu }\right) \right\vert \leq \frac{%
C_{5}}{\nu }\text{ as }\nu \rightarrow \infty ,\text{ }\left\vert D^{\alpha
}O\left( \frac{1}{\lambda }\right) \right\vert \leq \frac{C_{5}}{\lambda }%
\text{ as }\lambda \rightarrow \infty ,\text{ }\left\vert \alpha \right\vert
\leq 1.\text{ }  \label{8.05}
\end{equation}%
For the convenience of this proof rewrite the operator $L_{0}$ in (\ref{2.30}%
) as%
\begin{equation}
L_{0}u=\sum\limits_{i,j=1}^{n}a_{i,j}\left( x,t\right) u_{x_{i}x_{j}},\text{
}\left( x,t\right) \in Q_{T}.  \label{8.6}
\end{equation}%
Next, by (\ref{2.6})%
\begin{equation}
a_{i,j}\in C^{1}\left( \overline{Q}_{T}\right) ,\text{ }\max_{i,j}\left\Vert
a_{i,j}\right\Vert _{C^{1}\left( \overline{Q}_{T}\right) }\leq B.
\label{8.7}
\end{equation}%
And (\ref{2.5}) becomes%
\begin{equation}
\left. 
\begin{array}{c}
\mu \left\vert \xi \right\vert ^{2}\leq
\sum\limits_{i,j=1}^{n}a_{i,j}\left( x,t\right) \xi _{i}\xi _{j},\text{ }%
\forall \xi \in \mathbb{R}^{n},\text{ }\forall \left( x,t\right) \in Q_{T},
\\ 
a_{i,j}\left( x,t\right) =a_{j,i}\left( x,t\right) .%
\end{array}%
\right.  \label{8.8}
\end{equation}

Denote $w=ue^{\lambda \psi _{2}^{-\nu }}.$ Then%
\begin{equation}
u=we^{-\lambda \psi _{2}^{-\nu }}.  \label{8.9}
\end{equation}%
Using (\ref{7.80}) and (\ref{8.9}), express derivatives of the function $%
u\left( x,t\right) $ via derivatives of the function $w\left( x,t\right) .$
We have%
\begin{equation}
\left. 
\begin{array}{c}
u_{t}=\left( w_{t}+\lambda \nu \psi _{2}^{-\nu -1}w\left( t-T/2\right)
/T^{2}\right) e^{-\lambda \psi _{2}^{-\nu }}, \\ 
u_{x_{1}}=\left( w_{x_{1}}+\lambda \nu \psi _{2}^{-\nu -1}w\right)
e^{-\lambda \psi _{2}^{-\nu }}, \\ 
u_{x_{1}x_{1}}=\left( w_{x_{1}x_{1}}+2\lambda \nu \psi _{2}^{-\nu
-1}w_{x_{1}}+\lambda ^{2}\nu ^{2}\psi _{2}^{-2\nu -2}\left( 1+O\left(
1/\lambda \right) \right) w\right) e^{-\lambda \psi _{2}^{-\nu }}, \\ 
u_{x_{1}x_{i}}=\left( w_{x_{1}x_{i}}+\lambda \nu \psi _{2}^{-\nu
-1}w_{x_{i}}\right) e^{-\lambda \psi _{2}^{-\nu }},\text{ }i\neq 1, \\ 
u_{x_{i}x_{j}}=w_{x_{i}x_{j}}e^{-\lambda \psi _{2}^{-\nu }},\text{ }i\neq 1%
\text{ and }j\neq 1.%
\end{array}%
\right.  \label{8.10}
\end{equation}%
Using (\ref{8.6}), (\ref{8.8}) and (\ref{8.10}), consider $\left(
u_{t}-L_{0}u\right) ^{2}\psi _{2}^{\nu +1}\varphi _{2,\lambda ,\nu },$ 
\begin{equation}
\left. 
\begin{array}{c}
\left( u_{t}-L_{0}u\right) ^{2}\psi _{2}^{\nu +2}\varphi _{2,\lambda ,\nu }=
\\ 
=\left( 
\begin{array}{c}
w_{t}-\sum\limits_{i,j=1}^{n}a_{i,j}w_{x_{i}x_{j}}-2\lambda \nu \psi
_{2}^{-\nu -1}\sum\limits_{i=1}^{n}a_{1,i}w_{x_{i}}- \\ 
-\lambda ^{2}\nu ^{2}\psi _{2}^{-2\nu -2}\left( 1+O\left( 1/\lambda \right)
\right) w%
\end{array}%
\right) ^{2}\psi _{2}^{\nu +2}.%
\end{array}%
\right.  \label{8.11}
\end{equation}%
We took into account here that by the second line of (\ref{8.8}) $%
a_{1,i}=a_{i,1}.$ Denote%
\begin{equation}
\left. 
\begin{array}{c}
z_{1}=w_{t}, \\ 
z_{2}=-\sum\limits_{i,j=1}^{n}a_{i,j}w_{x_{i}x_{j}}, \\ 
z_{3}=-2\lambda \nu \psi _{2}^{-\nu
-1}\sum\limits_{i=1}^{n}a_{1,i}w_{x_{i}}, \\ 
z_{4}=-\lambda ^{2}\nu ^{2}\psi _{2}^{-2\nu -2}\left( 1+O\left( 1/\lambda
\right) \right) w.%
\end{array}%
\right.  \label{8.13}
\end{equation}%
By (\ref{8.11})-(\ref{8.13})%
\begin{equation}
\left. 
\begin{array}{c}
\left( u_{t}-L_{0}u\right) ^{2}\psi _{2}^{\nu +2}\varphi _{2,\lambda ,\nu
}=\left( z_{1}+z_{2}+z_{3}+z_{4}\right) ^{2}\psi _{2}^{\nu +2}\geq \\ 
=\left( z_{1}+z_{3}\right) ^{2}\psi _{2}^{\nu +2}+2\left( z_{1}+z_{3}\right)
\left( z_{2}+z_{4}\right) \psi _{2}^{\nu +2}= \\ 
=\left( z_{1}^{2}+z_{3}^{2}+2z_{1}z_{3}+2z_{1}z_{2}\right) \psi _{2}^{\nu
+2}+ \\ 
+2z_{1}z_{4}\psi _{2}^{\nu +2}+2z_{2}z_{3}\psi _{2}^{\nu +2}+2z_{3}z_{4}\psi
_{2}^{\nu +2}.%
\end{array}%
\right.  \label{8.16}
\end{equation}

\textbf{Step 1}. Estimate from the below the term $2z_{1}z_{2}\psi _{2}^{\nu
+2}$ in (\ref{8.16}). By the second line of (\ref{8.8}) and (\ref{8.13})%
\begin{equation}
\left. 
\begin{array}{c}
2z_{1}z_{2}\psi _{2}^{\nu
+2}=-2\sum\limits_{i,j=1}^{n}a_{i,j}w_{x_{i}x_{j}}w_{t}\psi _{2}^{\nu +2}=
\\ 
=\sum\limits_{i,j=1}^{n}\left(
-a_{i,j}w_{x_{i}x_{j}}w_{t}-a_{i,j}w_{x_{j}x_{i}}w_{t}\right) \psi _{2}^{\nu
+2}= \\ 
=\sum\limits_{i,j=1}^{n}\left[ \left( -a_{i,j}w_{x_{i}}w_{t}\psi _{2}^{\nu
+2}\right) _{x_{j}}+\left( -a_{i,j}w_{x_{j}}w_{t}\psi _{2}^{\nu +2}\right)
_{x_{i}}\right] + \\ 
+\sum\limits_{i,j=1}^{n}a_{i,j}\left(
w_{x_{i}}w_{tx_{j}}+w_{x_{j}}w_{tx_{i}}\right) \psi _{2}^{\nu +2}+ \\ 
+\sum\limits_{i,j=1}^{n}\left( \left( a_{i,j}\psi _{2}^{\nu +2}\right)
_{x_{j}}w_{x_{i}}w_{t}+\left( a_{i,j}\psi _{2}^{\nu +2}\right)
_{x_{i}}w_{x_{j}}w_{t}\right) = \\ 
=\sum\limits_{i,j=1}^{n}\left( a_{i,j}w_{x_{i}}w_{x_{j}}\psi _{2}^{\nu
+2}\right) _{t}-\sum\limits_{i,j=1}^{n}\left( a_{i,j}\psi _{2}^{\nu
+2}\right) _{t}w_{x_{i}}w_{x_{j}}+ \\ 
+2z_{1}\sum\limits_{i,j=1}^{n}\left( a_{i,j}\psi _{2}^{\nu +2}\right)
_{x_{j}}w_{x_{i}}- \\ 
-\sum\limits_{i,j=1}^{n}\left( a_{i,j}\psi _{2}^{\nu +2}\right)
_{t}w_{x_{i}}w_{x_{j}}+\operatorname{div}\widetilde{U}_{1}+\partial _{t}\widetilde{V}%
_{1}.%
\end{array}%
\right.  \label{8.160}
\end{equation}%
Thus, using (\ref{8.7}) and (\ref{8.160}), we obtain 
\begin{equation}
\left. 
\begin{array}{c}
2z_{1}z_{2}\psi _{2}^{\nu +2}\geq -C_{5}\left\vert \nabla w\right\vert
^{2}+2z_{1}\sum\limits_{i,j=1}^{n}\left( a_{i,j}\psi _{2}^{\nu +2}\right)
_{x_{j}}w_{x_{i}}+ \\ 
+\operatorname{div}\widetilde{U}_{1}+\partial _{t}\widetilde{V}_{1},%
\end{array}%
\right.  \label{8.17}
\end{equation}%
\begin{equation}
\left. 
\begin{array}{c}
\operatorname{div}\widetilde{U}_{1}=\sum\limits_{i,j=1}^{n}\left[ \left(
-a_{i,j}w_{x_{i}}w_{t}\psi _{2}^{\nu +2}\right) _{x_{j}}+\left(
-a_{i,j}w_{x_{j}}w_{t}\psi _{2}^{\nu +2}\right) _{x_{i}}\right] , \\ 
\partial _{t}\widetilde{V}_{1}=\sum\limits_{i,j=1}^{n}\left(
a_{i,j}w_{x_{i}}w_{x_{j}}\psi _{2}^{\nu +2}\right) _{t}.%
\end{array}%
\right.  \label{8.18}
\end{equation}%
Using (\ref{7.1}), (\ref{7.100}), (\ref{7.81}) and (\ref{8.7}), we obtain%
\begin{equation}
-\sum\limits_{i,j=1}^{n}\left( a_{i,j}\psi _{2}^{\nu +2}\right)
_{t}w_{x_{i}}w_{x_{j}}\geq -C_{5}\nu \left\vert \nabla w\right\vert ^{2}.
\label{8.20}
\end{equation}%
Now, by (\ref{7.80})%
\begin{equation*}
\partial _{x_{j}}\psi _{2}^{\nu +2}=\left\{ 
\begin{array}{c}
\left( \nu +2\right) \psi _{2}^{\nu +1}\text{ if }j=1, \\ 
0\text{ if }j\neq 1.%
\end{array}%
\right.
\end{equation*}%
Hence, using (\ref{8.05}), we obtain%
\begin{equation}
2z_{1}\sum\limits_{i,j=1}^{n}\left( a_{i,j}\psi _{2}^{\nu +2}\right)
_{x_{j}}w_{x_{i}}=2z_{1}\left( \nu +2\right) \psi _{2}^{\nu
+2}\sum\limits_{i=1}^{n}\left( a_{i,1}\psi _{2}^{-1}+O_{i}\left( \frac{1}{%
\nu }\right) \right) w_{x_{i}}.  \label{8.21}
\end{equation}%
Thus, (\ref{8.17}), (\ref{8.20}) and (\ref{8.21}) imply%
\begin{equation}
\left. 
\begin{array}{c}
2z_{1}z_{2}\psi _{2}^{\nu +2}\geq -C_{5}\nu \left\vert \nabla w\right\vert
^{2}+2z_{1}\left( \nu +2\right) \psi _{2}^{\nu
+2}\sum\limits_{i=1}^{n}\left( a_{i,1}\psi _{2}^{-1}+O_{i}\left( 1/\nu
\right) \right) w_{x_{i}}+ \\ 
+\operatorname{div}\widetilde{U}_{1}+\partial _{t}\widetilde{V}_{1},%
\end{array}%
\right.  \label{8.22}
\end{equation}%
where $\widetilde{U}_{1}$ and $\widetilde{V}_{1}$ are given in the third
line of (\ref{8.18}).

\textbf{Step 2.} Estimate from the below the term $\left(
z_{1}^{2}+z_{3}^{2}+2z_{1}z_{3}+2z_{1}z_{2}\right) \psi _{2}^{\nu +2}$ in (%
\ref{8.16}). Using (\ref{8.22}), we obtain 
\begin{equation}
\left. 
\begin{array}{c}
\left( z_{1}^{2}+z_{3}^{2}+2z_{1}z_{3}+2z_{1}z_{2}\right) \psi _{2}^{\nu
+2}\geq \\ 
\geq \left( z_{1}^{2}+z_{3}^{2}\right) \psi _{2}^{\nu +2}-C_{5}\nu
\left\vert \nabla w\right\vert ^{2}+ \\ 
+2z_{1}\left( \nu +2\right) \psi _{2}^{\nu +2}\left[ \sum\limits_{i=1}^{n}%
\left( a_{i,1}\psi _{2}^{-1}+O_{i}\left( 1/\nu \right) \right)
w_{x_{i}}+z_{3}/\left( \nu +2\right) \right] .%
\end{array}%
\right.  \label{8.23}
\end{equation}%
By Cauchy-Schwarz inequality 
\begin{equation}
\left. 
\begin{array}{c}
2z_{1}\left( \nu +2\right) \psi _{2}^{\nu +2}\left[ \sum\limits_{i=1}^{n}%
\left( a_{i,1}\psi _{2}^{-1}+O\left( 1/\nu \right) \right)
w_{x_{i}}+z_{3}/\left( \nu +2\right) \right] \geq \\ 
\geq -z_{1}^{2}\psi _{2}^{\nu +2}-\left( \nu +2\right) ^{2}\psi _{2}^{\nu
+2} \left[ \sum\limits_{i=1}^{n}\left( a_{i,1}\psi _{2}^{-1}+O_{i}\left(
1/\nu \right) \right) w_{x_{i}}+z_{3}/\left( \nu +2\right) \right] ^{2}= \\ 
=-\left( z_{1}^{2}+z_{3}^{2}\right) \psi _{2}^{\nu +2}-\left( \nu +2\right)
^{2}\psi _{2}^{\nu +2}\left[ \sum\limits_{i=1}^{n}\left( a_{i,1}\psi
_{2}^{-1}+O_{i}\left( 1/\nu \right) \right) w_{x_{i}}\right] ^{2}- \\ 
-2\left( \nu +2\right) \psi _{2}^{\nu +1}z_{3}\left(
\sum\limits_{i=1}^{n}\left( a_{i,1}+O\left( 1/\nu \right) \right)
w_{x_{i}}\right) .%
\end{array}%
\right.  \label{8.24}
\end{equation}%
Using the third line of (\ref{8.13}), we obtain for the last line of (\ref%
{8.24})%
\begin{equation}
\left. 
\begin{array}{c}
-2\left( \nu +2\right) \psi _{2}^{\nu +1}z_{3}\left(
\sum\limits_{i=1}^{n}\left( a_{i,1}+O_{i}\left( 1/\nu \right) \right)
w_{x_{i}}\right) = \\ 
=4\lambda \nu \left( \nu +2\right) \left(
\sum\limits_{i=1}^{n}a_{i,1}w_{x_{i}}\right) ^{2}+ \\ 
+4\lambda \nu \left( \nu +2\right) \sum\limits_{i,j=1}^{n}O_{i}\left( 1/\nu
\right) a_{1,j}w_{x_{i}}w_{x_{j}}\geq \\ 
\geq -C_{5}\lambda \nu \left\vert \nabla w\right\vert ^{2}.%
\end{array}%
\right.  \label{8.25}
\end{equation}%
Substituting (\ref{8.25}) in (\ref{8.24}), we obtain%
\begin{equation*}
\left. 
\begin{array}{c}
2z_{1}\left( \nu +2\right) \psi _{2}^{\nu +2}\left[ \sum\limits_{i=1}^{n}%
\left( a_{i,1}\psi _{2}^{-1}+O\left( 1/\nu \right) \right)
w_{x_{i}}+z_{3}/\left( \nu +2\right) \right] \geq \\ 
\geq -\left( z_{1}^{2}+z_{3}^{2}\right) \psi _{2}^{\nu +2}-C_{5}\lambda \nu
\left( \nabla w\right) ^{2}.%
\end{array}%
\right.
\end{equation*}%
Comparing this with (\ref{8.23}), we obtain%
\begin{equation}
\left( z_{1}^{2}+z_{3}^{2}+2z_{1}z_{3}+2z_{1}z_{2}\right) \psi _{2}^{\nu
+2}\geq -C_{5}\lambda \nu \left\vert \nabla w\right\vert ^{2}+\operatorname{div}%
\widetilde{U}_{1}+\partial _{t}\widetilde{V}_{1}.  \label{8.26}
\end{equation}

\textbf{Step 3.} Estimate from the below the term $2z_{1}z_{4}\psi _{2}^{\nu
+2}$ in the fourth line of (\ref{8.16}). Using (\ref{8.05}) and (\ref{8.13}%
), we obtain 
\begin{equation*}
\left. 
\begin{array}{c}
2z_{1}z_{4}\psi _{2}^{\nu +2}=-2\lambda ^{2}\nu ^{2}\psi _{2}^{-\nu }\left(
1+O\left( 1/\lambda \right) \right) ww_{t}\geq \\ 
\geq \left( -\lambda ^{2}\nu ^{2}\psi _{2}^{-\nu }\left( 1+O\left( 1/\lambda
\right) \right) w^{2}\right) _{t}-C_{5}\lambda ^{2}\nu ^{3}\psi _{2}^{-\nu
-1}w^{2}= \\ 
=-C_{5}\lambda ^{2}\nu ^{3}\psi _{2}^{-\nu -1}w^{2}+\partial _{t}\widetilde{V%
}_{2}.%
\end{array}%
\right.
\end{equation*}%
Thus, 
\begin{equation}
\left. 
\begin{array}{c}
2z_{1}z_{4}\psi _{2}^{\nu +2}\geq -C_{5}\lambda ^{2}\nu ^{3}\psi _{2}^{-\nu
-1}w^{2}+\partial _{t}\widetilde{V}_{2}, \\ 
\widetilde{V}_{2}=-\lambda ^{2}\nu ^{2}\psi _{2}^{-\nu }\left( 1+O\left(
1/\lambda \right) \right) w^{2}.%
\end{array}%
\right.  \label{8.27}
\end{equation}

\textbf{Step 4.} Estimate from the below the term $2z_{2}z_{3}\psi _{2}^{\nu
+2}$ in the fourth line of (\ref{8.16}). Using the second line of (\ref{8.8}%
) and (\ref{8.13}), we obtain%
\begin{equation}
2z_{2}z_{3}\psi _{2}^{\nu +2}=2\lambda \nu \sum\limits_{i,j,k=1}^{n}\psi
_{2}a_{i,j}a_{1,k}\left(
w_{x_{i}x_{j}}w_{x_{k}}+w_{x_{j}x_{i}}w_{x_{k}}\right) .  \label{8.28}
\end{equation}%
We have 
\begin{equation*}
\left. 
\begin{array}{c}
w_{x_{i}x_{j}}w_{x_{k}}+w_{x_{j}x_{i}}w_{x_{k}}=\left(
w_{x_{i}}w_{x_{k}}\right) _{x_{j}}-w_{x_{i}}w_{x_{k}x_{j}}+\left(
w_{x_{j}}w_{x_{k}}\right) _{x_{i}}-w_{x_{j}}w_{x_{k}x_{i}}= \\ 
=\left( w_{x_{i}}w_{x_{k}}\right) _{x_{j}}+\left( w_{x_{j}}w_{x_{k}}\right)
_{x_{i}}+\left( w_{x_{i}}w_{x_{j}}\right) _{x_{k}}.%
\end{array}%
\right.
\end{equation*}%
Hence,%
\begin{equation*}
\left. 
\begin{array}{c}
\psi _{2}a_{i,j}a_{1,k}\left(
w_{x_{i}x_{j}}w_{x_{k}}+w_{x_{j}x_{i}}w_{x_{k}}\right) = \\ 
=\left( \psi _{2}a_{i,j}a_{1,k}w_{x_{i}}w_{x_{k}}\right) _{x_{j}}-\left(
\psi _{2}a_{i,j}a_{1,k}\right) _{x_{j}}w_{x_{i}}w_{x_{k}}+ \\ 
+\left( \psi _{2}a_{i,j}a_{1,k}w_{x_{j}}w_{x_{k}}\right) _{x_{i}}-\left(
\psi _{2}a_{i,j}a_{1,k}\right) _{x_{i}}w_{x_{j}}w_{x_{k}}+ \\ 
+\left( \psi _{2}a_{i,j}a_{1,k}w_{x_{i}}w_{x_{j}}\right) _{x_{k}}-\left(
\psi _{2}a_{i,j}a_{1,k}\right) _{x_{k}}w_{x_{i}}w_{x_{j}}.%
\end{array}%
\right.
\end{equation*}%
Hence, using (\ref{8.28}), we obtain 
\begin{equation}
2z_{2}z_{3}\psi _{2}^{\nu +2}\geq -C_{5}\lambda \nu \left\vert \nabla
w\right\vert ^{2}+\operatorname{div}\widetilde{U}_{2},  \label{8.29}
\end{equation}%
\begin{equation}
\left. \operatorname{div}\widetilde{U}_{2}=\sum\limits_{i,j,k=1}^{n}\left[ 
\begin{array}{c}
\left( \psi _{2}a_{i,j}a_{1,k}w_{x_{i}}w_{x_{k}}\right) _{x_{j}}+\left( \psi
_{2}a_{i,j}a_{1,k}w_{x_{j}}w_{x_{k}}\right) _{x_{i}}+ \\ 
+\left( \psi _{2}a_{i,j}a_{1,k}w_{x_{i}}w_{x_{j}}\right) _{x_{k}}%
\end{array}%
\right] .\right.  \label{8.30}
\end{equation}

\textbf{Step 5.} Estimate from the below the term $2z_{3}z_{4}\psi _{2}^{\nu
+2}$ in in the fourth line of (\ref{8.16}). Using (\ref{8.05}) and (\ref%
{8.13}), we obtain%
\begin{equation}
\left. 
\begin{array}{c}
2z_{3}z_{4}\psi _{2}^{\nu +2}=4\lambda ^{3}\nu ^{3}\psi _{2}^{-2\nu
-1}\left( 1+O\left( 1/\lambda \right) \right)
\sum\limits_{i=1}^{n}a_{1,i}w_{x_{i}}w= \\ 
=4\lambda ^{3}\nu ^{3}\psi _{2}^{-2\nu -1}\left( 1+O\left( 1/\lambda \right)
\right) a_{11}w_{x_{1}}w+ \\ 
+4\lambda ^{3}\nu ^{3}\psi _{2}^{-2\nu -1}\left( 1+O\left( 1/\lambda \right)
\right) \sum\limits_{i=2}^{n}a_{1,i}w_{x_{i}}w= \\ 
=\left( 2\lambda ^{3}\nu ^{3}\psi _{2}^{-2\nu -1}\left( 1+O\left( 1/\lambda
\right) \right) a_{1,1}w^{2}\right) _{x_{1}}+ \\ 
\_+2\lambda ^{3}\nu ^{3}\left( 2\nu +1\right) \psi _{2}^{-2\nu -2}\left(
1+O\left( 1/\lambda \right) \right) a_{1,1}\left( 1+O\left( 1/\nu \right)
\right) w^{2}+ \\ 
+\sum\limits_{i=2}^{n}\left( 2\lambda ^{3}\nu ^{3}\psi _{2}^{-2\nu
-1}\left( 1+O\left( 1/\lambda \right) \right) a_{1,i}w^{2}\right) _{x_{i}}-
\\ 
-2\lambda ^{3}\nu ^{3}\psi _{2}^{-2\nu -1}\left( 1+O\left( 1/\lambda \right)
\right) \sum\limits_{i=2}^{n}\left( a_{1,i}\right) _{x_{i}}w^{2}\geq \\ 
\geq C_{5}\lambda ^{3}\nu ^{4}\psi _{2}^{-2\nu -2}w^{2}+\operatorname{div}\widetilde{%
U}_{3}.%
\end{array}%
\right.  \label{8.300}
\end{equation}%
In (\ref{8.300}) we have used the inequality $a_{11}\geq \mu ,$ which
follows from (\ref{8.8}). Thus, 
\begin{equation}
2z_{3}z_{4}\psi _{2}^{\nu +2}\geq C_{5}\lambda ^{3}\nu ^{4}\psi _{2}^{-2\nu
-2}w^{2}+\operatorname{div}\widetilde{U}_{3},  \label{8.31}
\end{equation}%
\begin{equation}
\left. \operatorname{div}\widetilde{U}_{3}=\sum\limits_{i=1}^{n}\left( 2\lambda
^{3}\nu ^{3}\psi _{2}^{-2\nu -1}\left( 1+O\left( 1/\lambda \right) \right)
a_{1,i}w^{2}\right) _{x_{i}}.\right.  \label{8.32}
\end{equation}

\textbf{Step 6.} Sum up estimates (\ref{8.26}), (\ref{8.27}), (\ref{8.29})
and (\ref{8.31}). When doing so, take into account formulas (\ref{8.18}), (%
\ref{8.30}) and (\ref{8.32}) containing terms with $\operatorname{div}$ and $%
\partial _{t}.$ Furthermore, replace the function $w$ with the function $u$
using formula (\ref{8.9}). Using (\ref{8.16}), we obtain then%
\begin{equation}
\left. 
\begin{array}{c}
\left( u_{t}-L_{0}u\right) ^{2}\psi _{2}^{\nu +2}\varphi _{2,\lambda ,\nu
}\geq -C_{5}\lambda \nu \left\vert \nabla u\right\vert ^{2}\varphi
_{2,\lambda ,\nu }+C_{5}\lambda ^{3}\nu ^{4}\psi _{2}^{-2\nu -2}u^{2}\varphi
_{2,\lambda ,\nu }+ \\ 
+\operatorname{div}\widetilde{U}_{4}+\partial _{t}\widetilde{V}_{4}, \\ 
\forall \nu \geq \nu _{1},\text{ }\forall \lambda \geq \lambda _{1},\text{ }%
\forall u\in C^{2,1}\left( \overline{G\left( \gamma ,T\right) }\right) ,%
\end{array}%
\right.  \label{8.33}
\end{equation}%
\begin{equation}
\left\vert \widetilde{U}_{4}\right\vert +\left\vert \widetilde{V}%
_{4}\right\vert \leq C_{5}\lambda ^{3}\nu ^{3}\psi _{2}^{-2\nu -2}\left(
\left\vert \nabla u\right\vert ^{2}+u_{t}^{2}+u^{2}\right) \varphi
_{2,\lambda ,\nu }.  \label{8.34}
\end{equation}

An inconvenient property of estimate (\ref{8.33}) is the presence of the
non-positive term $-C_{5}\lambda \nu \left\vert \nabla u\right\vert
^{2}\varphi _{2,\lambda ,\nu }.$ Hence, we perform Step 7.

\textbf{Step 7. }Estimate from the below the term $\left(
u_{t}-L_{0}u\right) u\varphi _{2,\lambda ,\nu }.$ Using (\ref{7.80}), (\ref%
{8.6})-(\ref{8.8}) and Cauchy-Schwarz inequality, we obtain 
\begin{equation*}
\left. 
\begin{array}{c}
\left( u_{t}-L_{0}u\right) u\varphi _{2,\lambda ,\nu }\geq \left( \left(
u^{2}/2\right) \varphi _{2,\lambda ,\nu }\right)
_{t}+\sum\limits_{i,j=1}^{n}\left( -a_{i,j}u_{x_{j}}u\varphi _{2,\lambda
,\nu }\right) _{x_{j}}+ \\ 
+\sum\limits_{i,j=1}^{n}a_{i,j}u_{x_{j}}u_{x_{i}}\varphi _{2,\lambda ,\nu
}-\left( \mu /2\right) \left\vert \nabla u\right\vert ^{2}\varphi
_{2,\lambda ,\nu }-C_{5}\lambda ^{2}\nu ^{2}\psi _{2}^{-2\nu -2}u^{2}\varphi
_{2,\lambda ,\nu }\geq \\ 
\geq \left( \mu /2\right) \left\vert \nabla u\right\vert ^{2}\varphi
_{2,\lambda ,\nu }-C_{5}\lambda ^{2}\nu ^{2}\psi _{2}^{-2\nu -2}u^{2}\varphi
_{2,\lambda ,\nu }+ \\ 
+\left( u^{2}\varphi _{2,\lambda ,\nu }/2\right)
_{t}+\sum\limits_{i,j=1}^{n}\left( -a_{i,j}u_{x_{j}}u\varphi _{2,\lambda
,\nu }\right) _{x_{j}}.%
\end{array}%
\right.
\end{equation*}%
Thus,%
\begin{equation}
\left. 
\begin{array}{c}
\left( u_{t}-L_{0}u\right) u\varphi _{2,\lambda ,\nu }\geq \left( \mu
/2\right) \left\vert \nabla u\right\vert ^{2}\varphi _{2,\lambda ,\nu
}-C_{5}\lambda ^{2}\nu ^{2}\psi _{2}^{-2\nu -2}u^{2}\varphi _{2,\lambda ,\nu
}+ \\ 
+\left( \left( u^{2}/2\right) \varphi _{2,\lambda ,\nu }\right)
_{t}+\sum\limits_{i,j=1}^{n}\left( -a_{i,j}u_{x_{j}}u\varphi _{2,\lambda
,\nu }\right) _{x_{j}}.%
\end{array}%
\right.  \label{8.35}
\end{equation}

\textbf{Step 8. }Let $\eta >0$ be such a number that $\eta \mu >4$. Multiply
both sides of inequality (\ref{8.35}) by $C_{5}\lambda \nu \eta $ and sum up
with (\ref{8.33}). \ Since $\lambda ^{3}\nu ^{4}\psi _{2}^{-2\nu
-2}>>\lambda ^{3}\nu ^{3}\psi _{2}^{-2\nu -2}$ for $\nu \geq \nu _{1}$ with
sufficiently large $\nu _{1},$ then we obtain using (\ref{8.34}) 
\begin{equation}
\left. 
\begin{array}{c}
C_{5}\lambda \nu \eta \left( u_{t}-L_{0}u\right) u\varphi _{2,\lambda ,\nu
}+\left( u_{t}-L_{0}u\right) ^{2}\psi _{2}^{\nu +2}\varphi _{2,\lambda ,\nu
}\geq \\ 
\geq C_{5}\lambda \nu \left\vert \nabla u\right\vert ^{2}\varphi _{2,\lambda
,\nu }+C_{5}\lambda ^{3}\nu ^{4}\psi _{2}^{-2\nu -2}u^{2}\varphi _{2,\lambda
,\nu }+\operatorname{div}\widetilde{U}+\partial _{t}\widetilde{V}, \\ 
\left\vert \widetilde{U}\right\vert +\left\vert \widetilde{V}\right\vert
\leq C_{5}\lambda ^{3}\nu ^{3}\psi _{2}^{-2\nu -2}\left( \left\vert \nabla
u\right\vert ^{2}+u_{t}^{2}+u^{2}\right) \varphi _{2,\lambda ,\nu }, \\ 
\forall \nu \geq \nu _{1},\text{ }\forall \lambda \geq \lambda _{1},\text{ }%
\forall u\in C^{2,1}\left( \overline{G\left( \gamma ,T\right) }\right) .%
\end{array}%
\right.  \label{8.36}
\end{equation}%
The target estimate (\ref{8.4}) of this theorem can be obtained immediately
from (\ref{8.36}) if applying Cauchy-Schwarz inequality to the first term in
the first line of (\ref{8.36}). \ \ $\square $

\subsection{The third Carleman estimate}

\label{sec:8.3}

Let $T_{1}>0$ be a number and $\lambda \geq 1$ be a large parameter, which
we will choose later. Denote%
\begin{equation}
Q_{T_{1}}=\Omega \times \left( 0,T_{1}\right) ,\text{ }S_{T_{1}}=\partial
\Omega \times \left( 0,T_{1}\right) .  \label{8.37}
\end{equation}
Denote%
\begin{equation}
H_{0}^{2}\left( Q_{T_{1}}\right) =\left\{ u\in H^{2}\left( Q_{T_{1}}\right)
:u\mid _{S_{T_{1}}}=0\right\} .  \label{8.38}
\end{equation}

\textbf{Theorem 8.3} (the third Carleman estimate). \emph{Assume that
conditions (\ref{8.6})-(\ref{8.8}) are satisfied with the replacement of }$%
Q_{T}$\emph{\ with the time cylinder }$Q_{T_{1}}$\emph{\ defined in (\ref%
{8.37}). Then there exists a sufficiently large number }$\lambda
_{2}=\lambda _{2}\left( \mu ,B,Q_{T_{1}}\right) \geq 1$\emph{\ and a number }%
$C_{6}=C_{6}\left( \mu ,B,Q_{T_{1}}\right) >0,$\emph{\ both numbers
depending only on listed parameters, such that the following Carleman
estimate holds}%
\begin{equation}
\left. 
\begin{array}{c}
\int\limits_{Q_{T_{1}}}\left( u_{t}-L_{0}u\right) ^{2}\exp \left( 2\left(
t+1\right) ^{\lambda }\right) dxdt\geq \\ 
\geq C_{6}\sqrt{\lambda }\int\limits_{Q_{T_{1}}}\left( \nabla u\right)
^{2}\exp \left( 2\left( t+1\right) ^{\lambda }\right) dxdt+ \\ 
+C_{6}\lambda ^{2}\int\limits_{Q_{T_{1}}}u^{2}\exp \left( 2\left(
t+1\right) ^{\lambda }\right) dxdt- \\ 
-\exp \left( 3\left( T_{1}+1\right) ^{\lambda }\right) \left\Vert u\left(
x,T_{1}\right) \right\Vert _{L_{2}\left( \Omega \right)
}^{2}-C_{6}\left\Vert \left\vert \nabla u\left( x,0\right) \right\vert
\right\Vert _{L_{2}\left( \Omega \right) }^{2}+ \\ 
+\left( 1/3\right) \lambda e^{2}\left\Vert u\left( x,0\right) \right\Vert
_{L_{2}\left( \Omega \right) }^{2}, \\ 
\forall \lambda \geq \lambda _{2},\text{ }\forall u\in H_{0}^{2}\left(
Q_{T_{1}}\right) .%
\end{array}%
\right.  \label{8.39}
\end{equation}

\textbf{Remark 8.1.} \emph{Both a novel and delicate point of estimate (\ref%
{8.39}) is the presence of the non-negative term }$\left( 1/3\right) \lambda
e^{2}\left\Vert u\left( x,0\right) \right\Vert _{L_{2}\left( \Omega \right)
}^{2}$\emph{\ in it. It is exactly this term, which will allow us to obtain
the logarithmic stability estimate (\ref{7.14}) in the course of the proof
of Theorem 7.1 in subsection 10.1. Indeed, while a similar Carleman estimate
was proven in \cite{KY}, this term is absent there. As a result, an analog
of (\ref{7.14}) was not obtained in \cite{KY}. The latter is unlike our
by-product results in section 11.}

\textbf{Proof of Theorem 8.3.} We prove this theorem in five steps. In this
proof, $C_{6}=C_{6}\left( \mu ,B,Q_{T_{1}}\right) >0$ denotes different
positive numbers depending only on listed parameters. Introduce the new
function 
\begin{equation}
w\left( x,t\right) =u\left( x,t\right) \exp \left( \left( t+1\right)
^{\lambda }\right) .  \label{8.40}
\end{equation}%
Then 
\begin{equation}
\left. 
\begin{array}{c}
u=w\exp \left( -\left( t+1\right) ^{\lambda }\right) , \\ 
u_{t}=\left( w_{t}-\lambda \left( t+1\right) ^{\lambda -1}w\right) \exp
\left( -\left( t+1\right) ^{\lambda }\right) , \\ 
u_{x_{i}x_{j}}=w_{x_{i}x_{j}}\exp \left( -\left( t+1\right) ^{\lambda
}\right) .%
\end{array}%
\right.  \label{8.040}
\end{equation}%
Hence, by (\ref{8.6})%
\begin{equation}
\left. 
\begin{array}{c}
\left( u_{t}-L_{0}u\right) ^{2}\exp \left( 2\left( t+1\right) ^{\lambda
}\right) = \\ 
=\left[ w_{t}-\left( \lambda \left( t+1\right) ^{\lambda
-1}w+\sum\limits_{i,j=1}^{n}a_{ij}w_{x_{i}x_{j}}\right) \right] ^{2}\geq \\ 
\geq w_{t}^{2}-2w_{t}\left( \lambda \left( t+1\right) ^{\lambda
-1}w+\sum\limits_{i,j=1}^{n}a_{ij}w_{x_{i}x_{j}}\right) .%
\end{array}%
\right.  \label{8.41}
\end{equation}

\textbf{Step 1.} Estimate $-2\lambda \left( t+1\right) ^{\lambda -1}w_{t}w.$
We have%
\begin{equation}
\left. 
\begin{array}{c}
-2\lambda \left( t+1\right) ^{\lambda -1}w_{t}w=\left( -\lambda \left(
t+1\right) ^{\lambda -1}w^{2}\right) _{t}+\lambda \left( \lambda -1\right)
\left( t+1\right) ^{\lambda -2}w^{2}\geq \\ 
\geq \left( -\lambda \left( t+1\right) ^{\lambda -1}w^{2}\right)
_{t}+C_{6}\lambda ^{2}\left( t+1\right) ^{\lambda }w^{2}.%
\end{array}%
\right.  \label{8.42}
\end{equation}

\textbf{Step 2}. Estimate 
\begin{equation*}
-2w_{t}\sum\limits_{i,j=1}^{n}a_{ij}w_{x_{i}x_{j}}.
\end{equation*}%
Using $a_{ij}=a_{ji},$ we obtain 
\begin{equation*}
\left. 
\begin{array}{c}
-2w_{t}\sum\limits_{i,j=1}^{n}a_{ij}w_{x_{i}x_{j}}= \\ 
=\sum\limits_{i,j=1}^{n}\left( -2a_{ij}w_{t}w_{x_{i}}\right)
_{x_{j}}+\sum\limits_{i,j=1}^{n}a_{ij}\left(
w_{tx_{j}}w_{x_{i}}+w_{tx_{i}}w_{x_{j}}\right)
+\sum\limits_{i,j=1}^{n}\left( a_{ij}\right) _{x_{j}}w_{t}w_{x_{i}}= \\ 
=\sum\limits_{i,j=1}^{n}\left( -2a_{ij}w_{t}w_{x_{i}}\right)
_{x_{j}}+\sum\limits_{i,j=1}^{n}a_{ij}\left( w_{x_{i}}w_{x_{j}}\right)
_{t}+\sum\limits_{i,j=1}^{n}\left( a_{ij}\right) _{x_{j}}w_{t}w_{x_{i}}= \\ 
=\sum\limits_{i,j=1}^{n}\left( -2a_{ij}w_{t}w_{x_{i}}\right)
_{x_{j}}+\sum\limits_{i,j=1}^{n}\left( a_{ij}\right)
_{x_{j}}w_{t}w_{x_{i}}+\sum\limits_{i,j=1}^{n}\left(
a_{ij}w_{x_{i}}w_{x_{j}}\right) _{t}-\sum\limits_{i,j=1}^{n}\left(
a_{ij}\right) _{t}w_{x_{i}}w_{x_{j}}\geq \\ 
\geq -w_{t}^{2}/2-C_{6}\left( \nabla w\right)
^{2}+\sum\limits_{i,j=1}^{n}\left( -2a_{ij}w_{t}w_{x_{i}}\right)
_{x_{j}}+\sum\limits_{i,j=1}^{n}\left( a_{ij}w_{x_{i}}w_{x_{j}}\right) _{t}.%
\end{array}%
\right.
\end{equation*}%
Thus, we have proven that 
\begin{equation*}
-2w_{t}\sum\limits_{i,j=1}^{n}a_{ij}w_{x_{i}x_{j}}\geq -\frac{w_{t}^{2}}{2}%
-C_{6}\left( \nabla w\right) ^{2}+\sum\limits_{i,j=1}^{n}\left(
-2a_{ij}w_{t}w_{x_{i}}\right) _{x_{j}}+\sum\limits_{i,j=1}^{n}\left(
a_{ij}w_{x_{i}}w_{x_{j}}\right) _{t}.
\end{equation*}%
Summing this up with (\ref{8.42}) and taking into account (\ref{8.40})-(\ref%
{8.41}), we obtain%
\begin{equation}
\left. 
\begin{array}{c}
\left( u_{t}-L_{0}u\right) ^{2}\exp \left( 2\left( t+1\right) ^{\lambda
}\right) \geq C_{6}\lambda ^{2}\left( t+1\right) ^{\lambda }u^{2}\exp \left(
2\left( t+1\right) ^{\lambda }\right) - \\ 
-C_{6}\left( \nabla u\right) ^{2}\exp \left( 2\left( t+1\right) ^{\lambda
}\right) + \\ 
+\left( -\lambda \left( t+1\right) ^{\lambda -1}u^{2}\exp \left( 2\left(
t+1\right) ^{\lambda }\right) \right) _{t}+ \\ 
+\left( \sum\limits_{i,j=1}^{n}\left( a_{ij}u_{x_{i}}u_{x_{j}}\exp \left(
2\left( t+1\right) ^{\lambda }\right) \right) \right)
_{t}+\sum\limits_{i,j=1}^{n}\left( -2a_{ij}w_{t}w_{x_{i}}\right) _{x_{j}}.%
\end{array}%
\right.  \label{8.43}
\end{equation}

An inconvenient point of estimate (\ref{8.43}) is the presence of the
non-positive term

$-C_{6}\left( \nabla u\right) ^{2}\exp \left( 2\left( t+1\right) ^{\lambda
}\right) $ in it. Hence, we use Step 3.

\textbf{Step 3. }Estimate $\left( u_{t}-L_{0}u\right) u\exp \left( 2\left(
t+1\right) ^{\lambda }\right) .$ Using Cauchy-Schwarz inequality (\ref{8.7})
and (\ref{8.8}), we obtain%
\begin{equation*}
\left. 
\begin{array}{c}
\left( u_{t}-L_{0}u\right) u\exp \left( 2\left( t+1\right) ^{\lambda
}\right) =\left( u_{t}-\sum\limits_{i,j=1}^{n}a_{ij}u_{x_{i}x_{j}}\right)
u\exp \left( 2\left( t+1\right) ^{\lambda }\right) = \\ 
=\left( u^{2}\exp \left( 2\left( t+1\right) ^{\lambda }\right) /2\right)
_{t}-\lambda \left( t+1\right) ^{\lambda -1}u^{2}\exp \left( 2\left(
t+1\right) ^{\lambda }\right) + \\ 
+\left( -\sum\limits_{i,j=1}^{n}a_{ij}u_{x_{i}}u\exp \left( 2\left(
t+1\right) ^{\lambda }\right) \right)
_{x_{j}}+\sum\limits_{i,j=1}^{n}a_{ij}u_{x_{i}}u_{x_{j}}\exp \left( 2\left(
t+1\right) ^{\lambda }\right) + \\ 
+\sum\limits_{i,j=1}^{n}\left( a_{ij}\right) _{x_{j}}u_{x_{i}}u\exp \left(
2\left( t+1\right) ^{\lambda }\right) \geq \\ 
\geq \left( \mu /2\right) \left( \nabla u\right) ^{2}\exp \left( 2\left(
t+1\right) ^{\lambda }\right) -C_{6}\lambda \left( t+1\right) ^{\lambda
}\exp \left( 2\left( t+1\right) ^{\lambda }\right) u^{2}+ \\ 
+\left( u^{2}\exp \left( 2\left( t+1\right) ^{\lambda }\right) /2\right)
_{t}+\left( -\sum\limits_{i,j=1}^{n}a_{ij}u_{x_{i}}u\exp \left( 2\left(
t+1\right) ^{\lambda }\right) \right) _{x_{j}}.%
\end{array}%
\right.
\end{equation*}%
Thus, we have proven that 
\begin{equation}
\left. 
\begin{array}{c}
\left( u_{t}-L_{0}u\right) u\exp \left( 2\left( t+1\right) ^{\lambda
}\right) \geq \\ 
\geq \left( \mu /2\right) \left( \nabla u\right) ^{2}\exp \left( 2\left(
t+1\right) ^{\lambda }\right) -C_{6}\lambda \left( t+1\right) ^{\lambda
}\exp \left( 2\left( t+1\right) ^{\lambda }\right) u^{2}+ \\ 
+\left( u^{2}\exp \left( 2\left( t+1\right) ^{\lambda }\right) /2\right)
_{t}+\left( -\sum\limits_{i,j=1}^{n}a_{ij}u_{x_{i}}u\exp \left( 2\left(
t+1\right) ^{\lambda }\right) \right) _{x_{j}}.%
\end{array}%
\right.  \label{8.44}
\end{equation}

\textbf{Step 4.} Choose a sufficiently large number $\lambda _{2}\geq 1$
such that $\mu \sqrt{\lambda _{2}}/2>2C_{6}.$ Below in this proof $\lambda
\geq \lambda _{2}.$ Multiply both sides of (\ref{8.44}) by $\sqrt{\lambda }$
and sum up with (\ref{8.43}). We obtain%
\begin{equation}
\left. 
\begin{array}{c}
\left( u_{t}-L_{0}u\right) ^{2}\exp \left( 2\left( t+1\right) ^{\lambda
}\right) +\sqrt{\lambda }\left( u_{t}-L_{0}u\right) u\exp \left( 2\left(
t+1\right) ^{\lambda }\right) \geq \\ 
\geq C_{6}\sqrt{\lambda }\left( \nabla u\right) ^{2}\exp \left( 2\left(
t+1\right) ^{\lambda }\right) +C_{6}\lambda ^{2}u^{2}\exp \left( 2\left(
t+1\right) ^{\lambda }\right) + \\ 
+\left( -\lambda \left( t+1\right) ^{\lambda -1}u^{2}\exp \left( 2\left(
t+1\right) ^{\lambda }+\sqrt{\lambda }u^{2}\exp \left( 2\left( t+1\right)
^{\lambda }\right) /2\right) \right) _{t}+ \\ 
+\left( \sum\limits_{i,j=1}^{n}\left( a_{ij}u_{x_{i}}u_{x_{j}}\exp \left(
2\left( t+1\right) ^{\lambda }\right) \right) \right) _{t}+ \\ 
+\sum\limits_{i,j=1}^{n}\left( -2a_{ij}w_{t}w_{x_{i}}\right)
_{x_{j}}+\left( -\sqrt{\lambda }\sum\limits_{i,j=1}^{n}a_{ij}u_{x_{i}}u\exp
\left( 2\left( t+1\right) ^{\lambda }\right) \right) _{x_{j}}.%
\end{array}%
\right.  \label{8.45}
\end{equation}%
By Cauchy-Schwarz inequality%
\begin{equation*}
\left. 
\begin{array}{c}
\sqrt{\lambda }\left( u_{t}-L_{0}u\right) u\exp \left( 2\left( t+1\right)
^{\lambda }\right) \leq \\ 
\leq \left( 1/2\right) \left( u_{t}-L_{0}u\right) ^{2}\exp \left( 2\left(
t+1\right) ^{\lambda }\right) +\left( 1/2\right) \lambda u^{2}\exp \left(
2\left( t+1\right) ^{\lambda }\right) .%
\end{array}%
\right.
\end{equation*}%
Substituting this in (\ref{8.45}), we obtain%
\begin{equation}
\left. 
\begin{array}{c}
\left( u_{t}-L_{0}u\right) ^{2}\exp \left( 2\left( t+1\right) ^{\lambda
}\right) \geq \\ 
\geq C_{6}\sqrt{\lambda }\left( \nabla u\right) ^{2}\exp \left( 2\left(
t+1\right) ^{\lambda }\right) +C_{6}\lambda ^{2}u^{2}\exp \left( 2\left(
t+1\right) ^{\lambda }\right) + \\ 
+\left( -\left( 2/3\right) \lambda \left( t+1\right) ^{\lambda -1}u^{2}\exp
\left( 2\left( t+1\right) ^{\lambda }\right) \right) _{t}+ \\ 
+\left( \left( 2/3\right) \sqrt{\lambda }u^{2}\exp \left( 2\left( t+1\right)
^{\lambda }\right) /2\right) _{t}+ \\ 
+\left( \left( 2/3\right) \sum\limits_{i,j=1}^{n}\left(
a_{ij}u_{x_{i}}u_{x_{j}}\exp \left( 2\left( t+1\right) ^{\lambda }\right)
\right) \right) _{t}+ \\ 
+\sum\limits_{i,j=1}^{n}\left( -\left( 4/3\right)
a_{ij}w_{t}w_{x_{i}}\right) _{x_{j}}+\left( -\left( 2/3\right) \sqrt{\lambda 
}\sum\limits_{i,j=1}^{n}a_{ij}u_{x_{i}}u\exp \left( 2\left( t+1\right)
^{\lambda }\right) \right) _{x_{j}}.%
\end{array}%
\right.  \label{8.46}
\end{equation}

\textbf{Step 5. }Integrate (\ref{8.46}) over the time cylinder $Q_{T_{1}}$
using Gauss formula and (\ref{8.38}). As to the term in the fifth line of (%
\ref{8.46}), using (\ref{8.7}) and (\ref{8.8}), we obtain%
\begin{equation*}
\left. 
\begin{array}{c}
\int\limits_{Q_{T_{1}}}\left( \left( 2/3\right)
\sum\limits_{i,j=1}^{n}\left( a_{ij}u_{x_{i}}u_{x_{j}}\exp \left( 2\left(
t+1\right) ^{\lambda }\right) \right) \right) _{t}dxdt= \\ 
=\left( 2/3\right) \exp \left( 2\left( T_{1}+1\right) ^{\lambda }\right)
\int\limits_{\Omega }\left(
\sum\limits_{i,j=1}^{n}a_{ij}u_{x_{i}}u_{x_{j}}\right) \left(
x,T_{1}\right) dx- \\ 
-\left( 2/3\right) e^{2}\int\limits_{\Omega }\left(
\sum\limits_{i,j=1}^{n}a_{ij}u_{x_{i}}u_{x_{j}}\right) \left( x,0\right)
dx\geq \\ 
\geq \left( 2/3\right) \mu \exp \left( 2\left( T_{1}+1\right) ^{\lambda
}\right) \left\Vert \left\vert \nabla u\left( x,T_{1}\right) \right\vert
\right\Vert _{L_{2}\left( \Omega \right) }^{2}-C_{6}\left\Vert \left\vert
\nabla u\left( x,0\right) \right\vert \right\Vert _{L_{2}\left( \Omega
\right) }^{2}\geq \\ 
\geq -C_{6}\left\Vert \left\vert \nabla u\left( x,0\right) \right\vert
\right\Vert _{L_{2}\left( \Omega \right) }^{2}.%
\end{array}%
\right.
\end{equation*}%
Hence,%
\begin{equation}
\left. 
\begin{array}{c}
\int\limits_{Q_{T_{1}}}\left( u_{t}-L_{0}u\right) ^{2}\exp \left( 2\left(
t+1\right) ^{\lambda }\right) \geq \\ 
\geq C_{6}\int\limits_{Q_{T_{1}}}\left( \sqrt{\lambda }\left( \nabla
u\right) ^{2}+\lambda ^{2}u^{2}\right) \exp \left( 2\left( t+1\right)
^{\lambda }\right) dxdt+ \\ 
+\left( 2/3\right) \left( \lambda -\sqrt{\lambda }\right)
e^{2}\int\limits_{\Omega }u^{2}\left( x,0\right) dx-C_{6}\left\Vert
\left\vert \nabla u\left( x,0\right) \right\vert \right\Vert _{L_{2}\left(
\Omega \right) }^{2}- \\ 
-\left( 2/3\right) \left( \lambda \left( T_{1}+1\right) ^{\lambda -1}-\sqrt{%
\lambda }\right) \exp \left( 2\left( T_{1}+1\right) ^{\lambda }\right)
\int\limits_{\Omega }u^{2}\left( x,T_{1}\right) dx.%
\end{array}%
\right.  \label{8.47}
\end{equation}%
Next, $\left( \lambda \left( T_{1}+1\right) ^{\lambda -1}-\sqrt{\lambda }%
\right) \leq \lambda \left( T_{1}+1\right) ^{\lambda -1}.$ Also, since $%
\lambda $ is sufficiently large, then

$\lambda -\sqrt{\lambda }>\lambda /2.$ Hence, (\ref{8.47}) implies 
\begin{equation*}
\left. 
\begin{array}{c}
\int\limits_{Q_{T_{1}}}\left( u_{t}-L_{0}u\right) ^{2}\exp \left( 2\left(
t+1\right) ^{\lambda }\right) \geq \\ 
\geq C_{6}\int\limits_{Q_{T_{1}}}\left( \sqrt{\lambda }\left( \nabla
u\right) ^{2}+\lambda ^{2}u^{2}\right) \exp \left( 2\left( t+1\right)
^{\lambda }\right) dxdt- \\ 
-\left( 2/3\right) \lambda \left( T_{1}+1\right) ^{\lambda -1}\exp \left(
2\left( T_{1}+1\right) ^{\lambda }\right) \int\limits_{\Omega }u^{2}\left(
x,T_{1}\right) dx \\ 
+\left( 1/3\right) \lambda e^{2}\int\limits_{\Omega }u^{2}\left( x,0\right)
dx-C_{6}\left\Vert \left\vert \nabla u\left( x,0\right) \right\vert
\right\Vert _{L_{2}\left( \Omega \right) }^{2}.%
\end{array}%
\right.
\end{equation*}%
This estimate, in turn implies the target estimate (\ref{8.39}) of Theorem
8.3. \ $\square $

\section{Proof of Theorem 6.1}

\label{sec:9}

Let $H$ be a Hilbert space with the norm $\left\Vert \cdot \right\Vert _{H}.$
We denote $H_{N}=H\times H\times ...\times H,$ $N$ times. We define the norm
in $H_{N}$ as%
\begin{equation}
\left\Vert \left( u_{0},...,u_{N-1}\right) ^{T}\right\Vert
_{H_{N}}^{2}=\sum\limits_{j=0}^{N-1}\left\Vert u_{j}\right\Vert _{H}^{2},%
\text{ }\forall \left( u_{0},...,u_{N-1}\right) ^{T}\in H_{N}.  \label{9.1}
\end{equation}

Introduce the $N-$D vector function 
\begin{equation}
V\left( x,t\right) =\left( v_{0},...,v_{N-1}\right) ^{T}\left( x,t\right) ,%
\text{ }\left( x,t\right) \in Q_{T}.  \label{9.2}
\end{equation}%
Multiply equation (\ref{6.4}) sequentially by functions $\Psi _{k}\left(
x_{0}\right) ,$ $k=0,...,N-1$ and then integrate each of resulting equations
with respect to $x_{0}\in \left( a,b\right) .$ Thus, we obtain a system of $%
N $ coupled parabolic equations with respect to the components of the vector
function $V\left( x,t\right) $ in (\ref{9.1}). The principal part of the PDE
operator of this system is $W_{N}\left( V_{t}-L_{0}V\right) ,$ where the $%
N\times N$ matrix $W_{N}$ was defined via (\ref{4.01})-(\ref{4.3}). By
Theorem 4.1 the matrix $W_{N}$ is invertible. Hence, we multiply this system
by $W_{N}^{-1}.$ Next, it is convenient for the further proof to use (\ref%
{5.18})-(\ref{5.191}) in order to rewrite the resulting system in a more
general form as the following inequality%
\begin{equation}
\left\vert V_{t}-L_{0}V\right\vert \left( x,t\right) \leq \overline{C}%
_{N}\left( \left\vert \nabla V\right\vert +\left\vert V\right\vert \right)
\left( x,t\right) ,\text{ }\left( x,t\right) \in Q_{T},  \label{9.3}
\end{equation}%
Here and below $\overline{C}_{N}>0$ denotes different numbers depending on
the same parameters as the number $\overline{C}$ in (\ref{5.190}) as well as
on $N$. It follows from the fifth line of (\ref{6.5}), (\ref{6.15}) and (\ref%
{9.2}) that%
\begin{equation}
\left\Vert V\mid _{\Gamma _{T}}\right\Vert _{H_{N}^{1,1}\left( \Gamma
_{T}\right) }\leq \delta ,\text{ }\left\Vert \partial _{n}V\mid _{\Gamma
_{T}}\right\Vert _{L_{2,N}\left( \Gamma _{T}\right) }\leq \delta .
\label{9.4}
\end{equation}

Let $\varphi _{1,\lambda ,\nu _{2}}\left( x,t\right) $ be the function
defined in (\ref{8.1}), where the number $\nu _{2}\geq \nu _{0},$ and $\nu
_{0}$ is the same as in (\ref{8.100}). We will choose the number $\nu _{2}$
later. Square both sides of (\ref{9.3}), then multiply the resulting
inequality by $\varphi _{1,\lambda ,\nu _{2}}\left( x,t\right) ,$ use
Cauchy-Schwarz inequality and the first Carleman estimate (\ref{8.2}). We
obtain 
\begin{equation}
\left. 
\begin{array}{c}
\overline{C}_{N}\left( \left\vert \nabla V\right\vert ^{2}+\left\vert
V\right\vert ^{2}\right) \varphi _{1,\lambda ,\nu _{2}}\geq \\ 
\geq C_{4}\lambda \left\vert \nabla V\right\vert ^{2}\varphi _{1,\lambda
,\nu _{2}}+C_{4}\lambda ^{3}\left\vert V\right\vert ^{2}\varphi _{1,\lambda
,\nu _{2}}+\operatorname{div}U+W_{t}, \\ 
\left\vert U\right\vert +\left\vert W\right\vert \leq C_{4}\lambda
^{3}\left( \left\vert \nabla V\right\vert ^{2}+V_{t}^{2}+V^{2}\right)
\varphi _{1,\lambda ,\nu _{2}}, \\ 
\forall \left( x,t\right) \in \overline{Q_{T}\left( \gamma _{1},\gamma
_{2},t_{0}\right) },\text{ }\forall \lambda \geq \lambda _{0},%
\end{array}%
\right.  \label{9.5}
\end{equation}%
where the number $\lambda _{0}$ was defined in (\ref{8.100}). Choose the
number $\lambda ^{\left( 1\right) }\geq \lambda _{0}$ so large that $%
C_{4}\lambda ^{\left( 1\right) }\geq 2\overline{C}_{N}.$ Below in this proof 
$\lambda \geq \lambda ^{\left( 1\right) }.$ Using (\ref{9.5}), we obtain%
\begin{equation}
\left. 
\begin{array}{c}
\widetilde{C}\lambda \left( \left\vert \nabla V\right\vert ^{2}+\left\vert
V\right\vert ^{2}\right) \varphi _{1,\lambda ,\nu _{2}}\leq -\left( \func{div%
}U+W_{t}\right) , \\ 
\forall \left( x,t\right) \in \overline{Q_{T}\left( \gamma _{1},\gamma
_{2},t_{0}\right) },\text{ }\forall \lambda \geq \lambda ^{\left( 1\right) }.%
\end{array}%
\right.  \label{9.6}
\end{equation}%
Here and below $\widetilde{C}>0$ denotes different numbers depending on the
combined set of parameters for $C_{4}$ and $\overline{C}_{N}.$

It follows from (\ref{6.9})-(\ref{6.13}) and (\ref{8.1}) that 
\begin{equation}
\left. 
\begin{array}{c}
\varphi _{1,\lambda ,\nu _{2}}\left( x,t\right) =\exp \left( 2\lambda \gamma
_{2}^{-\nu _{2}}\right) ,\text{ }\forall \left( x,t\right) \in \partial
_{1}Q_{T}\left( \gamma _{1},\gamma _{2},t_{0}\right) , \\ 
\max_{\overline{\partial _{2}Q_{T}\left( \gamma _{1},\gamma
_{2},t_{0}\right) }}\varphi _{1,\lambda ,\nu _{2}}\left( x,t\right) =\exp
\left( 2\lambda \gamma _{1}^{-\nu _{2}}\right) , \\ 
\min_{\overline{Q_{T}^{\omega }\left( \gamma _{1},\gamma _{2},t_{0}\right) }%
}\varphi _{1,\lambda ,\nu _{2}}\left( x,t\right) \geq \exp \left( 2\lambda
\left( \gamma _{2}-\omega \right) ^{-\nu _{2}}\right)%
\end{array}%
\right.  \label{9.7}
\end{equation}%
Integrating inequality (\ref{9.6}) over the domain $Q_{T}\left( \gamma
_{1},\gamma _{2},t_{0}\right) $ and using Gauss formula, (\ref{6.120})-(\ref%
{6.13}), the third line of (\ref{9.5}) and (\ref{9.7}), we obtain%
\begin{equation}
\left. 
\begin{array}{c}
\widetilde{C}\lambda ^{2}\exp \left( 2\lambda \gamma _{2}^{-\nu _{2}}\right)
\int\limits_{\partial _{1}Q_{T}\left( \gamma _{1},\gamma _{2},t_{0}\right)
}\left( \left\vert \nabla V\right\vert ^{2}+V_{t}^{2}+V^{2}\right) dS+ \\ 
+\widetilde{C}\lambda ^{2}\exp \left( 2\lambda \gamma _{1}^{-\nu
_{2}}\right) \int\limits_{\partial _{2}Q_{T}\left( \gamma _{1},\gamma
_{2},t_{0}\right) }\left( \left\vert \nabla V\right\vert
^{2}+V_{t}^{2}+V^{2}\right) dS\geq \\ 
\geq \int\limits_{Q_{T}\left( \gamma _{1},\gamma _{2},t_{0}\right) }\left(
\left\vert \nabla V\right\vert ^{2}+\left\vert V\right\vert ^{2}\right)
\varphi _{1,\lambda ,\nu _{1}}dxdt\geq \\ 
\geq \int\limits_{Q_{T}^{\omega }\left( \gamma _{1},\gamma
_{2},t_{0}\right) }\left( \left\vert \nabla V\right\vert ^{2}+\left\vert
V\right\vert ^{2}\right) \varphi _{1,\lambda ,\nu _{2}}dxdt\geq \\ 
\geq \exp \left( 2\lambda \left( \gamma _{2}-\omega \right) ^{-\nu
_{2}}\right) \int\limits_{Q_{T}^{\omega }\left( \gamma _{1},\gamma
_{2},t_{0}\right) }\left( \left\vert \nabla V\right\vert ^{2}+\left\vert
V\right\vert ^{2}\right) dxdt.%
\end{array}%
\right.  \label{9.8}
\end{equation}%
Using (\ref{6.12}), (\ref{6.14}), (\ref{6.15}) and (\ref{9.8}), we obtain 
\begin{equation}
\left. 
\begin{array}{c}
C_{1}\exp \left( 3\lambda \gamma _{1}^{-\nu _{2}}\right) \delta
^{2}+C_{1}\lambda ^{2}\exp \left( 2\lambda \gamma _{2}^{-\nu _{2}}\right)
\geq \\ 
\geq \exp \left( 2\lambda \left( \gamma _{2}-\omega \right) ^{-\nu
_{2}}\right) \int\limits_{Q_{T}^{\omega }\left( \gamma _{1},\gamma
_{2},t_{0}\right) }\left( \left\vert \nabla V\right\vert ^{2}+\left\vert
V\right\vert ^{2}\right) dxdt.%
\end{array}%
\right.  \label{9.9}
\end{equation}%
Divide both sides of (\ref{9.9}) by $\exp \left( 2\lambda \left( \gamma
_{2}-\omega \right) ^{-\nu _{2}}\right) .$ Then, we obtain for sufficiently
large $\nu _{2}\geq \nu _{0}$ such that $\left( \left( \gamma _{2}-\omega
\right) /\gamma _{2}\right) ^{\nu _{2}}<1/4$ 
\begin{equation}
\left. 
\begin{array}{c}
\int\limits_{Q_{T}^{\omega }\left( \gamma _{1},\gamma _{2},t_{0}\right)
}\left( \left\vert \nabla V\right\vert ^{2}+\left\vert V\right\vert
^{2}\right) dxdt\leq C_{1}\exp \left( 3\lambda \gamma _{1}^{-\nu
_{2}}\right) \delta ^{2}+ \\ 
+C_{1}\exp \left[ -\lambda \left( \gamma _{2}-\omega \right) ^{-\nu _{2}}%
\right] .%
\end{array}%
\right.  \label{9.14}
\end{equation}%
Choose $\lambda =\lambda \left( \delta \right) $ such that 
\begin{equation}
\lambda \left( \delta \right) \geq \lambda ^{\left( 1\right) }\text{ and }%
\exp \left( 3\lambda \gamma _{1}^{-\nu _{2}}\right) \delta ^{2}=\delta .
\label{9.15}
\end{equation}%
Hence, 
\begin{equation}
\lambda \left( \delta \right) =\ln \left[ \left( \frac{1}{\delta }\right)
^{\left( \gamma _{1}^{\nu _{2}}\right) /3}\right] .  \label{9.16}
\end{equation}%
The choice (\ref{9.15}), (\ref{9.16}) is possible if $\delta \in \left(
0,\delta _{1}\right) ,$ where $\delta _{1}\in \left( 0,1\right) $ is a
sufficiently small number depending on the same parameters as those listed
in (\ref{6.16}). For $\lambda \left( \delta \right) $ satisfying (\ref{9.16})%
\begin{equation}
\exp \left[ -\lambda \left( \gamma _{2}-\omega \right) ^{-\nu _{2}}\right]
=\delta ^{\left( \gamma _{1}/\left( \gamma _{2}-\omega \right) \right) ^{\nu
_{2}}/3}>\delta .  \label{9.17}
\end{equation}%
Denote 
\begin{equation}
\rho _{1}=\frac{1}{6}\left( \frac{\gamma _{1}}{\gamma _{2}-\omega }\right)
^{\nu _{2}}\in \left( 0,\frac{1}{6}\right) .  \label{9.18}
\end{equation}%
Then (\ref{9.1}) and (\ref{9.14})-(\ref{9.18}) imply%
\begin{equation}
\left\Vert V\right\Vert _{H_{N}^{1,0}\left( Q_{T}^{\omega }\left( \gamma
_{1},\gamma _{2},t_{0}\right) \right) }\leq C_{1}\delta ^{\rho _{1}},\text{ }%
\forall \delta \in \left( 0,\delta _{1}\right) .  \label{9.19}
\end{equation}

The first target estimate (\ref{6.17}) of this theorem follows immediately
from (\ref{9.1}), (\ref{9.2}) and (\ref{9.19}). Next, using (\ref{9.3}) and (%
\ref{9.19}), we obtain%
\begin{equation}
\left\Vert V_{t}-L_{0}V\right\Vert _{L_{2,N}\left( Q_{T}^{\omega }\left(
\gamma _{1},\gamma _{2},t_{0}\right) \right) }\leq C_{1}\delta ^{\rho _{1}},%
\text{ }\forall \delta \in \left( 0,\delta _{1}\right) .  \label{9.20}
\end{equation}%
Hence, using (\ref{5.18})-(\ref{5.191}), (\ref{6.3}) and (\ref{9.20}), we
obtain (\ref{6.18}), which is the second target estimate of this theorem. $\
\ \ \square $

\section{ Proofs of Theorems 7.1 and 7.2}

\label{sec:10}

\subsection{Proof of Theorem 7.1}

\label{sec:10.1}

We again use notations (\ref{9.1}) and (\ref{9.2}). Acting completely
similarly with the proof of Theorem 6.1, we obtain inequality (\ref{9.3}).
As to the lateral Cauchy data for the vector function $V\left( x,t\right) ,$
using the fifth line of (\ref{7.5}), (\ref{7.810}) and (\ref{9.2}) we obtain
the following analog of (\ref{9.4})%
\begin{equation}
\left\Vert V\mid _{S_{T}}\right\Vert _{H_{N}^{1,1}\left( S_{T}\right) }\leq
\delta ,\text{ }\left\Vert \partial _{n}V\mid _{S_{T,d_{2}}}\right\Vert
_{L_{2,N}\left( S_{T,d_{2}}\right) }\leq \delta .  \label{10.1}
\end{equation}

Consider the function $\varphi _{2,\lambda ,\nu _{3}}\left( x,t\right) $
defined in (\ref{8.3}), where the number $\nu _{3}\geq \nu _{1},$ and $\nu
_{1}$ is the same as in (\ref{8.04}). We will choose the number $\nu _{3}$
later. Square both sides of (\ref{9.3}), multiply then the resulting
inequality by $\varphi _{2,\lambda ,\nu _{3}}\left( x,t\right) ,$ use
Cauchy-Schwarz inequality and the second Carleman estimate (\ref{8.4}). Then
integrate the resulting inequality over the domain $G\left( \gamma ,T\right) 
$ using Gauss formula. We obtain 
\begin{equation}
\left. 
\begin{array}{c}
\overline{C}_{N}\int\limits_{G\left( \gamma ,T\right) }\left( \left\vert
\nabla V\right\vert ^{2}+V^{2}\right) \varphi _{2,\lambda ,\nu _{3}}dxdt\geq
\\ 
\geq C_{5}\int\limits_{G\left( \gamma ,T\right) }\lambda \left( \left\vert
\nabla V\right\vert ^{2}+V^{2}\right) \varphi _{2,\lambda ,\nu _{3}}dxdt- \\ 
-C_{5}\lambda ^{3}\int\limits_{\partial G\left( \gamma ,T\right) }\left(
\left\vert \nabla V\right\vert ^{2}+V_{t}^{2}+V^{2}\right) \varphi
_{2,\lambda ,\nu _{3}}dS,\text{ }\forall \lambda \geq \lambda _{1}.%
\end{array}%
\right.  \label{10.2}
\end{equation}%
Choose $\lambda _{1,1}\geq \lambda _{1}$ so large that $C_{5}\lambda _{1,1}>2%
\overline{C}_{N}.$ Then (\ref{8.4}) leads to%
\begin{equation}
\left. 
\begin{array}{c}
C_{2}\lambda ^{2}\int\limits_{\partial G\left( \gamma ,T\right) }\left(
\left\vert \nabla V\right\vert ^{2}+V_{t}^{2}+V^{2}\right) \varphi
_{2,\lambda ,\nu _{3}}dS\geq \\ 
\geq \int\limits_{G\left( \gamma ,T\right) }\left( \left\vert \nabla
V\right\vert ^{2}+V^{2}\right) \varphi _{2,\lambda ,\nu _{3}}dxdt,\text{ }%
\forall \lambda \geq \lambda _{1,1}.%
\end{array}%
\right.  \label{10.003}
\end{equation}%
Now, by (\ref{7.83}) the integral in the left hand side of (\ref{10.003})
can be written as: 
\begin{equation}
\left. 
\begin{array}{c}
\int\limits_{\partial G\left( \gamma ,T\right) }\left( \left\vert \nabla
V\right\vert ^{2}+V_{t}^{2}+V^{2}\right) \varphi _{2,\lambda ,\nu _{3}}dS=
\\ 
=\sum\limits_{k=1}^{4}\int\limits_{\partial _{k}G\left( \gamma ,T\right)
}\left( \left\vert \nabla V\right\vert ^{2}+V_{t}^{2}+V^{2}\right) \varphi
_{2,\lambda ,\nu _{3}}dS.%
\end{array}%
\right.  \label{10.004}
\end{equation}%
Estimate from the above each of integrals in the third line of (\ref{10.004}%
). Using (\ref{7.80}), (\ref{7.810}), (\ref{7.83}), (\ref{8.3}) and (\ref%
{10.1}), we obtain%
\begin{equation}
\sum\limits_{k=1}^{3}\int\limits_{\partial _{k}G\left( \gamma ,T\right)
}\left( \left\vert \nabla V\right\vert ^{2}+V_{t}^{2}+V^{2}\right) \varphi
_{2,\lambda ,\nu _{3}}dS\leq C_{2}\exp \left( 2\lambda \gamma ^{-\nu
_{3}}\right) \delta ^{2}.  \label{10.005}
\end{equation}%
Next, by (\ref{6.14}),\emph{\ }(\ref{7.80}), (\ref{7.83}), (\ref{8.3}) and (%
\ref{9.2}) 
\begin{equation}
\int\limits_{\partial _{4}G\left( \gamma ,T\right) }\left( \left\vert
\nabla V\right\vert ^{2}+V_{t}^{2}+V^{2}\right) \varphi _{2,\lambda ,\nu
_{3}}dS\leq C_{2}\exp \left( 2\lambda \left( 2\gamma +A\right) ^{-\nu
_{3}}\right) .  \label{10.006}
\end{equation}%
Since by (\ref{7.84}) $G^{\varepsilon }\left( \gamma ,T\right) \subset
G\left( \gamma ,T\right) $ and also $\varphi _{2,\lambda ,\nu _{3}}\left(
x,t\right) \geq \exp \left[ 2\lambda \left( 2\left( \gamma -\varepsilon
\right) +A\right) ^{-\nu _{3}}\right] $ in $G^{\varepsilon }\left( \gamma
,T\right) ,$ then (\ref{10.003})-(\ref{10.006}) lead to%
\begin{equation*}
\left. 
\begin{array}{c}
\exp \left[ 2\lambda \left( 2\left( \gamma -\varepsilon \right) +A\right)
^{-\nu _{3}}\right] \int\limits_{G^{\varepsilon }\left( \gamma ,T\right)
}\left( \left\vert \nabla V\right\vert ^{2}+V^{2}\right) dxdt\leq \\ 
\leq C_{2}\exp \left( 3\lambda \gamma ^{-\nu _{3}}\right) \delta
^{2}+C_{2}\exp \left( 2\lambda \left( 2\gamma +A\right) ^{-\nu _{3}}\right) .%
\end{array}%
\right.
\end{equation*}%
Divide both sides of this inequality by $\exp \left[ 2\lambda \left( 2\left(
\gamma -\varepsilon \right) +A\right) ^{-\nu _{3}}\right] .$ We obtain%
\begin{equation}
\left. 
\begin{array}{c}
\int\limits_{G^{\varepsilon }\left( \gamma ,T\right) }\left( \left\vert
\nabla V\right\vert ^{2}+V^{2}\right) dxdt\leq C_{2}\exp \left( 3\lambda
\gamma ^{-\nu _{3}}\right) \delta ^{2}+ \\ 
+C_{2}\exp \left[ -2\lambda \left( 2\left( \gamma -\varepsilon \right)
+A\right) ^{-\nu _{3}}\left( 1-\left( 2\left( \gamma -\varepsilon \right)
+A\right) /\left( \left( 2\gamma +A\right) \right) ^{\nu _{3}}\right) \right]
.%
\end{array}%
\right.  \label{10.007}
\end{equation}

Choosing $\nu _{3}$ so large that%
\begin{equation*}
\left( \frac{2\left( \gamma -\varepsilon \right) +A}{2\gamma +A}\right)
^{\nu _{3}}<\frac{1}{2},
\end{equation*}%
we obtain from (\ref{10.007}) 
\begin{equation}
\left. 
\begin{array}{c}
\int\limits_{G^{\varepsilon }\left( \gamma ,T\right) }\left( \left\vert
\nabla V\right\vert ^{2}+\left\vert V\right\vert ^{2}\right) dxdt\leq
C_{2}\exp \left( 3\lambda \gamma ^{-\nu _{3}}\right) \delta ^{2}+ \\ 
+C_{2}\exp \left[ -\lambda \left( 2\left( \gamma -\varepsilon \right)
+A\right) ^{-\nu _{3}}\right] ,\text{ }\forall \lambda \geq \lambda _{1,1}.%
\end{array}%
\right.  \label{10.008}
\end{equation}%
Next, proceeding completely similarly with the the part of the proof of
Theorem 6.1, which starts from (\ref{9.14}), we obtain from (\ref{10.008})
estimates (\ref{7.10}) and (\ref{7.11}), which are first two target
estimates of this theorem.

We now proceed to obtain the final estimate (\ref{7.14}) of this theorem.
Fix a point $x_{0}\in \left( a,b\right) .$ For brevity, we do not indicate
the dependence on $x_{0}$ in this proof below. Using (\ref{7.103}) and (\ref%
{7.84}), we obtain 
\begin{equation}
\Omega \times \left( t_{\varepsilon }^{-},\sigma \right) \subset \Omega
\times \left( t_{\varepsilon }^{-},t_{\varepsilon }^{+}\right) \subset
G^{\varepsilon }\left( \gamma ,T\right) .  \label{10.03}
\end{equation}%
Hence, the mean value theorem, (\ref{6.1}), (\ref{7.10}) and (\ref{9.2})
imply that there exists a number $\sigma _{1}\in \left( t_{\varepsilon
}^{-},\sigma \right) $ such that 
\begin{equation}
\left\Vert V\left( x,\sigma _{1}\right) \right\Vert _{H_{N}^{1,0}\left(
\Omega \right) }\leq \frac{1}{\sigma -t_{\varepsilon }^{-}}C_{2}\delta
^{\rho _{2}},\text{ }\forall \delta \in \left( 0,\delta _{2}\right) .
\label{10.3}
\end{equation}%
We now rewrite (\ref{9.3}) only for $\left( x,t\right) \in Q_{\sigma
_{1}}=\Omega \times \left( 0,\sigma _{1}\right) ,$%
\begin{equation}
\left\vert V_{t}-L_{0}V\right\vert \left( x,t\right) \leq \overline{C}%
_{N}\left( \left\vert \nabla V\right\vert +\left\vert V\right\vert \right)
\left( x,t\right) ,\text{ }\left( x,t\right) \in Q_{\sigma _{1}}.
\label{10.4}
\end{equation}%
By (\ref{7.110}) 
\begin{equation}
V\mid _{\partial \Omega \times \left( 0,\sigma _{1}\right) }=0.  \label{10.5}
\end{equation}%
Hence, $V\in H_{0,N}^{2}\left( Q_{\sigma _{1}}\right) ,$ see (\ref{8.38})
and (\ref{9.1}).

Square both sides of (\ref{10.4}). Then multiply the resulting equation by
the function $\exp \left( 2\left( t+1\right) ^{\lambda }\right) $ of Theorem
8.3, integrate over $Q_{\sigma _{1}},$ use Cauchy-Schwarz inequality, (\ref%
{8.38}), (\ref{10.5}) as well as Theorem 8.3. We obtain%
\begin{equation}
\left. 
\begin{array}{c}
\int\limits_{Q_{\sigma _{1}}}\left( \left\vert \nabla V\right\vert
^{2}+V^{2}\right) \exp \left( 2\left( t+1\right) ^{\lambda }\right) dxdt\geq
\\ 
\geq C_{3}\sqrt{\lambda }\int\limits_{Q_{\sigma _{1}}}\left\vert \nabla
V\right\vert ^{2}\exp \left( 2\left( t+1\right) ^{\lambda }\right) dxdt+ \\ 
+C_{3}\lambda ^{2}\int\limits_{Q_{\sigma _{1}}}V^{2}\exp \left( 2\left(
t+1\right) ^{\lambda }\right) dxdt- \\ 
-\exp \left( 3\left( T+1\right) ^{\lambda }\right) \left\Vert V\left(
x,\sigma _{1}\right) \right\Vert _{L_{2,N}\left( \Omega \right)
}^{2}-C_{3}\left\Vert \left\vert \nabla V\left( x,0\right) \right\vert
^{2}\right\Vert _{L_{2,N}\left( \Omega \right) }^{2}+ \\ 
+\left( 1/3\right) \lambda e^{2}\left\Vert V\left( x,0\right) \right\Vert
_{L_{2,N}\left( \Omega \right) }^{2},\text{ }\forall \lambda \geq \lambda
_{2}.%
\end{array}%
\right.  \label{10.7}
\end{equation}%
Choose a sufficiently large number%
\begin{equation*}
\lambda _{3}=\lambda _{3}\left( \Omega ,T,\varepsilon ,\gamma ,A,\mu
,N,\left( a,b\right) ,K,\overline{C},\sigma ,M\right) \geq \lambda _{2}
\end{equation*}%
depending only on listed parameters such that $C_{3}\sqrt{\lambda _{3}}\geq
2.$ Below in this proof $\lambda \geq \lambda _{3}.$ Thus, (\ref{10.3}) and (%
\ref{10.7}) imply%
\begin{equation}
\left. 
\begin{array}{c}
C_{3}\exp \left( 3\left( T+1\right) ^{\lambda }\right) \delta ^{2\rho
_{2}}+C_{3}\left\Vert \nabla V\left( x,0\right) \right\Vert _{L_{2,N}\left(
\Omega \right) }^{2}\geq \\ 
\geq C_{3}\sqrt{\lambda }\int\limits_{Q_{\sigma _{1}}}\left( \nabla
V\right) ^{2}\exp \left( 2\left( t+1\right) ^{\lambda }\right) dxdt+ \\ 
+C_{3}\lambda ^{2}\int\limits_{Q_{\sigma _{1}}}V^{2}\exp \left( 2\left(
t+1\right) ^{\lambda }\right) dxdt+ \\ 
+\left( 1/3\right) \lambda e^{2}\left\Vert V\left( x,0\right) \right\Vert
_{L_{2,N}\left( \Omega \right) }^{2},\text{ }\forall \lambda \geq \lambda
_{3}.%
\end{array}%
\right.  \label{10.70}
\end{equation}

We rewrite this inequality in a simplified form as%
\begin{equation}
\left\Vert V\left( x,0\right) \right\Vert _{L_{2,N}\left( \Omega \right)
}^{2}\leq C_{3}\exp \left( 3\left( T+1\right) ^{\lambda }\right) \delta
^{2\rho _{2}}+\frac{C_{3}}{\lambda }\left\Vert \nabla V\left( x,0\right)
\right\Vert _{L_{2,N}\left( \Omega \right) }^{2},\forall \lambda \geq
\lambda _{3}.  \label{10.8}
\end{equation}%
Recall that $C_{3}>0$ denotes different numbers depending on parameters
listed in (\ref{7.01}). Hence, it follows from (\ref{7.001}) and (\ref{9.2})
that (\ref{10.8}) can be rewritten as%
\begin{equation*}
\left\Vert V\left( x,0\right) \right\Vert _{L_{2,N}\left( \Omega \right)
}^{2}\leq C_{3}\exp \left( 3\left( T+1\right) ^{\lambda }\right) \delta
^{2\rho _{2}}+\frac{C_{3}}{\lambda },\text{ }\forall \lambda \geq \lambda
_{3}.
\end{equation*}%
Choose $\lambda =\lambda \left( \delta \right) $ such that%
\begin{equation}
\lambda \geq \lambda _{3}\text{ and }\exp \left( 3\left( T+1\right)
^{\lambda }\right) \delta ^{2\rho _{2}}=\delta ^{\rho _{2}}.  \label{10.9}
\end{equation}%
The choice is possible for all $\delta \in \left( 0,\delta _{3}\right) ,$
where $\delta _{3}\in \left( 0,1\right) $ is a sufficiently small number
depending on the same parameters as those listed in (\ref{7.01}). By (\ref%
{10.9})%
\begin{equation*}
\lambda =\lambda \left( \delta \right) =\ln \left[ \frac{\ln \left( \delta
^{-\rho _{2}}\right) }{\ln \left( 3\left( T+1\right) \right) }\right] ,\text{
}\forall \delta \in \left( 0,\delta _{3}\right) .
\end{equation*}%
Substituting this in (\ref{10.8}), we obtain%
\begin{equation}
\left\Vert V\left( x,0\right) \right\Vert _{L_{2,N}\left( \Omega \right)
}\leq \frac{C_{3}}{\sqrt{\ln \left[ \ln \left( 1/\delta \right) \right] }},%
\text{ }\forall \delta \in \left( 0,\delta _{3}\right) .  \label{10.90}
\end{equation}%
The third target estimate (\ref{7.14}) of this theorem follows immediately
from estimate (\ref{10.90}). $\ \square $

\subsection{Proof of Theorem 7.2}

\label{sec:10.2}

Fix a point $x_{0}\in \left( a,b\right) .$ Again, for brevity, we do not
indicate in this proof the dependence on $x_{0}$. Using (\ref{7.10}), (\ref%
{9.2}), (\ref{10.03}) and the mean value theorem, we obtain that there
exists a number 
\begin{equation}
\eta \in \left( t_{\varepsilon }^{-},t_{\varepsilon }^{+}\right)
\label{10.011}
\end{equation}%
such that 
\begin{equation}
\left\Vert V\left( x,\eta \right) \right\Vert _{L_{2,N}\left( \Omega \right)
}\leq \frac{C_{2}}{d_{\varepsilon }}\delta ^{\rho _{2}},\text{ }\forall
\delta \in \left( 0,\delta _{2}\right) ,  \label{10.11}
\end{equation}%
where $d_{\varepsilon }$ is the number defined in (\ref{7.104}). We have in
terms of the scalar product in $\mathbb{R}^{N}$%
\begin{equation}
\left. 
\begin{array}{c}
V\left( V_{t}-L_{0}V\right) = \\ 
=\left( V^{2}/2\right) _{t}-\sum\limits_{i,j=1}^{n}2a_{i,j}V_{x_{i}x_{j}}V=
\\ 
=\left( V^{2}/2\right) _{t}-\sum\limits_{i,j=1}^{n}\left(
a_{i,j}V_{x_{i}}V\right)
_{x_{j}}+\sum\limits_{i,j=1}^{n}a_{i,j}V_{x_{i}}V_{x_{j}}+\sum%
\limits_{i,j=1}^{n}\left( a_{i,j}\right) _{x_{j}}V_{x_{i}}V\geq \\ 
\geq \left( V^{2}/2\right) _{t}-\sum\limits_{i,j=1}^{n}\left(
a_{i,j}V_{x_{i}}V\right) _{x_{j}}+\left( \mu /2\right) \left( \nabla
V\right) ^{2}-C_{2}V^{2},%
\end{array}%
\right.  \label{10.12}
\end{equation}%
where we have used (\ref{8.7}), the first line of (\ref{8.8}) and
Cauchy-Schwarz inequality. Note that by (\ref{7.101}), (\ref{7.103}) and (%
\ref{10.011}) $d_{2}<t_{\varepsilon }^{-}<\eta .$ Also, since $\rho _{2}\in
\left( 0,1/6\right) $ and the number $\delta \in \left( 0,\delta _{2}\right) 
$ is sufficiently small, then $\delta ^{2}<\delta ^{2\rho _{2}}.$ Hence,
integrating (\ref{10.12}) over the domain $Q_{t,\eta }=\Omega \times \left(
\eta ,t\right) $ for an arbitrary $t\in \left( \eta ,T\right) $ and using
Gauss formula, (\ref{10.1}) and (\ref{10.11}), we obtain%
\begin{equation}
\left. 
\begin{array}{c}
\left( 1/2\right) \int\limits_{\Omega }V^{2}\left( x,t\right) dx+\left( \mu
/2\right) \int\limits_{0}^{t}\left( \int\limits_{\Omega }\left\vert \nabla
V\right\vert ^{2}\left( x,\tau \right) dx\right) d\tau \leq \\ 
\leq C_{2}\delta ^{2\rho _{2}}+\int\limits_{0}^{t}\left(
\int\limits_{\Omega }V\left( V_{t}-L_{0}V\right) \left( x,\tau \right)
dx\right) d\tau + \\ 
+C_{2}\int\limits_{0}^{t}\left( \int\limits_{\Omega }V^{2}\left( x,\tau
\right) dx\right) d\tau .%
\end{array}%
\right.  \label{10.13}
\end{equation}%
As stated in the beginning of subsection 10.1, inequality (\ref{9.3}) is
valid. Hence, combining (\ref{9.3}) with (\ref{10.13}) and using again
Cauchy-Schwarz inequality, we obtain 
\begin{equation}
\left. 
\begin{array}{c}
\int\limits_{\Omega }V^{2}\left( x,t\right) dx+\left( \mu /2\right)
\int\limits_{0}^{t}\left( \int\limits_{\Omega }\left\vert \nabla
V\right\vert ^{2}\left( x,\tau \right) dx\right) d\tau \leq \\ 
\leq C_{2}\int\limits_{0}^{t}\left( \int\limits_{\Omega }V^{2}\left(
x,\tau \right) dx\right) d\tau +C_{2}\delta ^{2\rho _{2}}.%
\end{array}%
\right.  \label{10.14}
\end{equation}%
Applying Gr\"{o}nwall's theorem to (\ref{10.14}) and combining it with (\ref%
{7.10}), (\ref{7.15})\ and (\ref{9.2}), we obtain (\ref{7.16}), which is the
first target estimate of this theorem. The second target estimate (\ref{7.17}%
) is obtained via a simple combination of (\ref{6.1}), (\ref{6.3}), (\ref%
{7.16}), (\ref{9.2}) and (\ref{9.3}). $\ \square $

\section{Two By-Products}

\label{sec:11}

In this section we formulate Theorems 11.1 and 11.2, which are by-products
of Theorem 7.1.

Let $\Omega \subset \mathbb{R}^{n}$ be an arbitrary bounded domain with the
piecewise smooth boundary $\partial \Omega .$ We keep notations (\ref{2.2})
for $Q_{T}$ and $S_{T}$ and replace $T_{1}$ with $T$ in (\ref{8.38}). Let $%
L_{0}$ be the elliptic operator (\ref{8.6}) with properties (\ref{8.7}), (%
\ref{8.8}). Slightly changing notation (\ref{2.3}), we define the elliptic
operator $L$ as%
\begin{equation}
Lu=L_{0}u+\sum\limits_{j=1}^{n}b_{j}\left( x,t\right) u_{x_{j}}+b_{0}\left(
x,t\right) u,\text{ }\left( x,t\right) \in Q_{T},  \label{11.1}
\end{equation}%
where functions $b_{j}\in C\left( \overline{Q}_{T}\right) ,$ $j=0,...,n.$ We
assume that the following analog of (\ref{8.7}) holds%
\begin{equation}
\left\Vert b_{j}\right\Vert _{C\left( \overline{Q}_{T}\right) }\leq B,\text{ 
}j=0,...,n.  \label{11.2}
\end{equation}

We consider two problems with time reversed data.

\textbf{Problem 1.} \emph{Let the function }$u\in H_{0}^{2}\left(
Q_{T}\right) $\emph{\ satisfies the following conditions:}%
\begin{equation}
u_{t}=Lu\text{ in }Q_{T},  \label{11.3}
\end{equation}%
\begin{equation}
u\left( x,T\right) =F_{1}\left( x\right) .  \label{11.4}
\end{equation}%
\emph{Estimate the initial condition} $u\left( x,0\right) $ \emph{via the
function }$F_{1}\left( x\right) .$

In addition, we consider a more general Problem 2.

\textbf{Problem 2.}\emph{\ Let the function }$u\in H_{0}^{2}\left(
Q_{T}\right) $\emph{\ satisfies the differential inequality}%
\begin{equation}
\left\vert u_{t}-L_{0}u\right\vert \leq M\left( \left\vert \nabla
u\right\vert +\left\vert u\right\vert \right) \text{ in }Q_{T},  \label{11.5}
\end{equation}%
\emph{where }$M>0$\emph{\ is a number. In addition, let }%
\begin{equation}
u\left( x,T\right) =F_{2}\left( x\right) .  \label{11.6}
\end{equation}%
\emph{Estimate the initial condition} $u\left( x,0\right) $ \emph{via the
function }$F_{2}\left( x\right) .$

\textbf{Theorem 11.1.} \emph{Assume that conditions (\ref{11.1})-(\ref{11.4}%
) hold. Let }$\left\Vert \left\vert \nabla u\left( x,0\right) \right\vert
\right\Vert _{L_{2}\left( \Omega \right) }\leq D_{1},$\emph{\ where }$%
D_{1}>0 $\emph{\ is a known number.} \emph{Then there exists a sufficiently
small number }$\delta _{0,1}=\delta _{0,1}\left( \mu ,B,Q_{T}\right) \in
\left( 0,1\right) $\emph{\ depending only on listed parameters such that if }%
$\left\Vert F_{1}\right\Vert _{L_{2}\left( \Omega \right) }\leq \delta ,$%
\emph{\ then the following estimate is valid: }%
\begin{equation*}
\left\Vert u\left( x,0\right) \right\Vert _{L_{2}\left( \Omega \right) }\leq 
\frac{M_{1}}{\sqrt{\ln \left[ \ln \left( 1/\delta \right) \right] }},\text{ }%
\forall \delta \in \left( 0,\delta _{0,1}\right) ,
\end{equation*}%
\emph{where the number }$M_{1}=M_{1}\left( \mu ,B,Q_{T},D_{1}\right) >0$%
\emph{\ depends only on listed parameters.}

\textbf{Theorem 11.2. }\emph{Assume that conditions (\ref{11.1}), (\ref{11.2}%
), (\ref{11.5}) and (\ref{11.6}) hold. Let }$\left\Vert \left\vert \nabla
u\left( x,0\right) \right\vert \right\Vert _{L_{2}\left( \Omega \right)
}\leq D_{2},$\emph{\ where }$D_{2}>0$ \emph{is a known number. Then there
exists a sufficiently small number }$\delta _{0,2}=\delta _{0,2}\left( \mu
,M,Q_{T}\right) \in \left( 0,1\right) $\emph{\ depending only on listed
parameters such that if }$\left\Vert F_{2}\right\Vert _{L_{2}\left( \Omega
\right) }\leq \delta ,$\emph{\ then the following estimate is valid: }%
\begin{equation*}
\left\Vert u\left( x,0\right) \right\Vert _{L_{2}\left( \Omega \right) }\leq 
\frac{M_{2}}{\sqrt{\ln \left[ \ln \left( 1/\delta \right) \right] }},\text{ }%
\forall \delta \in \left( 0,\delta _{0,2}\right) ,
\end{equation*}%
\emph{where the number }$M_{2}=M_{2}\left( \mu ,B,M,Q_{T},D_{2}\right) >0$%
\emph{\ depends only on listed parameters.}

We omit proofs of these two theorems, since they are very similar to the
part of the proof of Theorem 7.1, which starts from (\ref{10.4}).

\section{Replacement of $x_{0}$ With a Temporal Parameter $t^{0}$}

\label{sec:12}

It is assumed in Inverse Problems 1 and 2 that $x_{0}$ is a spatial
parameter being involved in the initial condition $f\left( x,x_{0}\right) .$
However, one can replace $x_{0}$ with a temporal parameter $t^{0}\in \left(
a,b\right) ,$ which is involved in one of boundary conditions. For example,
one can replace the function $g_{0}\left( x,t,x_{0}\right) $ in (\ref{2.11})
with the function $\widehat{g}_{0}\left( x,t,t^{0}\right) $. In this case
the unknown initial condition $f\left( x,x_{0}\right) $ in (\ref{2.10})
should be replaced with the unknown initial condition $\widehat{f}\left(
x\right) .$ Then close analogs of above results can be proven under the
condition that functions $\Psi _{j}\left( x_{0}\right) $ in (\ref{6.1})
would be replaced with $\Psi _{j}\left( t^{0}\right) ,$ $t^{0}\in \left(
a,b\right) .$ Proofs would be completely similar with the above ones.

\textbf{Declaration}

\textbf{Competing interests} The author has no competing interests to
declare that are relevant to the content of this article.

\textbf{Funding information. }This research was supported by the National
Science Foundation grant DMS 2436227.

\end{document}